\documentclass{article}[12pt]
\usepackage{oldlfont, hyperref, natbib, epsfig, amsmath, cmbright}

\newtheorem{theorem}{Theorem}[section]
\newtheorem{corollary}{Corollary}[section]
\newtheorem{proposition}{Proposition}[section]
\newtheorem{lemma}{Lemma}[section]

\newcommand{\bel}{{\rm Bel}}
\newcommand{\pl}{{\rm Pl}}

\newcommand{\hil}{{\mathcal H}}
\newcommand{\bases}{\mathsf{B}(\mathcal{H}^\alpha)}
\newcommand{\basess}{{\mathsf B}({\mathcal H}^{\alpha+1})}

\newcommand{\boxf}{\hfill$\square$}
\newcommand{\ch}{{\rm ch}}
\newcommand{\Pp}{{^+\!P}}
\newcommand{\Pm}{{^-\!P}}
\newcommand{\kri}{{\mathbf M}}
\newcommand{\krip}{{^+{\mathbf M}}}

\newcommand{\orthop}{{\mathcal L}_P}
\newcommand{\A}{{\mathcal A}}
\newcommand{\supp}{{\rm supp}}
\newcommand{\onehalf}{\mbox{\small  1/2}}

\usepackage{amssymb}

\begin{document}


  \title{Modal logic approach to preferred bases in the quantum universe}
  \author{Andreas Martin Lisewski\thanks{Address: Department of Molecular and Human Genetics, Baylor College of Medicine, One Baylor Plaza, Houston, TX 77030, USA. Email: {\tt lisewski@bcm.tmc.edu}}}
  \date{31 March 2004}

\maketitle\thispagestyle{empty}

{\tableofcontents\thispagestyle{empty}}
\thispagestyle{empty}
\vfill
\pagebreak
  \begin{abstract}
We present a modal logic based approach to the so-called endophysical quantum universe. In particular, we treat the problem of preferred bases and that of state reduction by employing an eclectic collection of methods including Baltag's analytic non-wellfounded set theory, a modal logic interpretation of Dempster-Shafer theory, and results from the theory of isometric embeddings of discrete metrics. Two basic principles, the bisimulation principle and the principle of imperfection, are derived that permit us to conduct an inductive proof showing that a preferred basis emerges at each evolutionary stage of the quantum universe. These principles are understood as theoretical realizations of the paradigm according to which the physical universe is a simulation on a quantum computer and a second paradigm saying that physical degrees of freedom are a model of Poincar\'e's physical continuum. Several comments are given related to communication theory, to evolutionary biology, and to quantum gravity.  
  \end{abstract}
\pagebreak
\setcounter{page}{1}
\section{\label{Introduction}Introduction}

The present work aims to give an attempt for a finite and complete quantum description of the physical universe by using elements of modal logic and set theory. The term {\it finite} means that it is a system of only finitely many physical degrees of freedom. Completeness means that the universe is understood as a closed quantum system without an external classical world and without any observers outside the quantum system. Since all relevant physical phenomena have to be explained from {\it within} the quantum system, complete descriptions are often attributed as {\it endophysical}.  Hence, the apparent reality of a classical physical world along with our experienced reality of ourselves as observers with a free will becomes an {\it emergent} concept in this understanding. Emergence is seen as a phenomenon known from physical systems with a sufficient number of physical constituents and with a sufficiently complex evolution of the latter. Emergence often manifests itself in global physical behavior that cannot be understood properly by looking only at the system's local constituents.\smallskip 

The idea of treating the universe  as a closed and discrete quantum system is not new. For example, already in 1982 Feynman \citep{fey1982} explored some implications of the assumption that the universe (the {\it physical world} in Feynman's terms) is a {\it quantum computer} representable as a tensor product of many finite Hilbert spaces of low dimension such as qubits. Considerations like this one suggest to look at the evolution of the universe as a run of a quantum automaton. This happens in a very similar manner to classical cellular automata-- like Conway's {\it Life} \citep{gar1970}, for example-- which evolve in a sequence of discrete steps. A quantum automaton is thus any finite system in some predefined initial state together with some rules that govern its step-wise evolution. The rules themselves have to be consistent with the laws of quantum physics; classical cellular automata can therefore only be a limiting case of quantum automata. But even in simple classical cellular automata rich varieties of complex patterns emerge \citep{wol2002}, thus it is at least not implausible to think of our physical universe as the output of a quantum automaton.\smallskip

In a recent series of papers \citep{ej2002, ej2003, ej2004a} Eakins and Jaroszkiewicz present this idea again using a more involved physical framework. Briefly stated, their approach postulates that the universe is represented in a Hilbert space $\hil$ of very large but finite and fixed dimensionality $N'$. A certain state vector $\Psi_\alpha \in \hil$ represents the current pure state of the universe. This state is an element of an orthonormal basis given through the family of non-degenerate eigenstates of a Hermitian operator $\Sigma^\alpha$ acting on $\hil$; the family of eigenstates of $\Sigma^\alpha$ is the so-called {\it preferred basis} while the operator itself is phrased the {\it self-test} of the universe. Also, they suggest that the state of the universe is subject to change.  This change is governed by some rules, yet unknown, that map $\Psi_\alpha$ onto its successor $\Psi_{\alpha+1}$. Thus the index $\alpha$ labels the successive stages of the universe and is called the {\it exotime}. These rules guarantee further that $\Psi_{\alpha+1}$ also is an eigenvector but this time of a different Hermitian operator $\Sigma^{\alpha+1}$. The main argument of Eakins and Jaroszkiewicz is that at each stage $\alpha$ the Hilbert space $\hil$ factors in a tensor product 
$$
\hil = \hil_1 \otimes \ldots \otimes \hil_{N}
$$
of $N$ Hilbert spaces $\hil_i$ each having a prime dimensionality $d_i = \dim \hil_i$. States in $\hil$ can be total factor states, they can also be totally entangled, and also it is possible to have states that contain factors of entangled states. Thus $\Psi_\alpha$ admits the general form
$$
\Psi_\alpha = \Psi_\alpha^1 \otimes \ldots \otimes \Psi_\alpha^{f_\alpha}
$$
where $f_\alpha \leq N$. Since it is believed that any self-test has the capacity to change  the factor structure of a given state $\Psi_\alpha$ when going from one stage $\alpha$ to the next stage $\alpha+1$, the corresponding transition amplitude calculated with Born's rule may or may not factorize. This observation allows to look at groups of factors that become entangled in the successor stage or at entangled states from subregisters that become factorized within the next stage. When followed over several successive stages, the transition amplitudes between states resemble the structure of causal sets (for details, see \citep{ej2003}); it is in this manner that the building blocks of Einstein locality seem to be accessible. Moreover, Eakins and Jaroszkiewicz speculate about further implications of their approach, such as the possibility that highly factorized states should correspond to a quantum system with emergent classical behavior.\medskip 

Although the quoted results and ideas surely motivate for further work into this direction, our immediate goal is to take a step back and to recapitulate the common assumptions and prerequisites that form the basis of this approach to a complete quantum universe. In doing so we list a group of questions that are at the source of all arguments presented in this work. 
\begin{enumerate}
\item If the universe admits a representation by means of a Hilbert space $\hil$ of fixed finite dimension $N'$, what causes the choice of the number $N'$? For now, there does not seem to be an immediate physical reason behind the choice of $N'$. We know that at present time this number must be gigantic but has this been the case throughout the history of the universe? In other words, is it necessary that the Hilbert space is static with a fixed number of dimensions?\footnote{For the approach of Eakins and Jaroszkiewicz this question implies another argument, namely, that $N'$ must not be prime since otherwise no non-trivial tensor product of subregisters is available. Is there a physical reason behind this?}  

\item We know that Hermitian operators represent observables in quantum physics, but why should the self-test of the universe $\Sigma^\alpha$ be Hermitian and non-degenerate. At least such an extrapolation from local physical experience to the universe as a whole is relatively bold. Must we simply accept it as a matter of fact or can we possibly find a reason that explains these properties of $\Sigma^\alpha$?

\item How does the preferred basis, i.e. the family of orthonormal eigenvectors of $\Sigma^\alpha$, emerge at each stage of the universe? This question-- also referred to as the problem of pointer states-- was recognized by Eakins and Jaroszkiewicz and has been a central issue in various approaches  to the measurement process in quantum physics (e.g., in the decoherence framework and in the  many-worlds approach).

\item How does state reduction or, more appropriately phrased, state {\it selection} occur at each stage of the universe. In \citep{ej2003} it has been plausibly argued that $\Psi_\alpha$ always is a pure state, but how does the universe make a choice between the available elements of the preferred basis? This question addresses the second central (and still undecided) issue of the measurement process. To put it differently, is von Neumann's formal characterization of the measurement procedure,  that is the distinction between processes of type II (unitary dynamics) and type I (reduction), the {\it final word} or can we do better in characterizing a (non-deterministic) process responsible for state selection?

\item What is the mathematical structure of the Hilbert space $\hil$? Since any quantum theory of the universe should propose an explicit construction of the Hilbert space of physical quantum states, we ask whether we can identify the preferred basis elements of $\hil$. This is the step of going from an abstract Hilbert space to a concrete Hilbert space where physical states are explicitly given.
 
\end{enumerate} 

Our intention is to discuss these five problems and to investigate possible solutions. All five problem statements are ultimately related to questions about the physical nature of observers who conduct measurements, about the observable objects, and about the process of measurement. Our strategy is to introduce two basic principles and to find their mathematical representations in order to gain further insights into these problem statements and, eventually, into the nature of the quantum universe. To the knowledge of the author, these principles as well as the mathematical methods related to them have not yet been widely used in this problem domain.\medskip

The first principle concerns our ability to perceive nature through experiments (every measurement we call an experiment). By experiments we do not only mean an experimental physical set-up and its conduction   in the usual sense but also the ultimate class of experiments that we carry out on ourselves in order to become aware of any experiment whatsoever, namely, our sensory perception.  The {\it imperfection principle} says that every experiment in nature has to be blurred in some sense. This means that there must always be a set of several measurement outcomes such that each   member of this set must not be perceptually separated from any other member of the same set. Experiments of this kind we call {\it imperfect experiments}, and hence the principle demands that any experiment in nature must be imperfect. This makes sense in many cases because empirically we know that experimental data has limited precision. But there are types of experiments where it is apparently more difficult to recognize the validity of the imperfection principle. For example, consider a Stern-Gerlach experiment with a detector screen placed behind the magnetic field. The spin value in $z$-direction of a spin $\onehalf$ particle shall be determined. Imagine the measurement outcome now is a dot at the upper half of the screen signalizing that the measured particle has a value of $+\onehalf$ in $z$-direction. As it seems, there is no fuzziness in the measurement outcome since the particle spin in $z$-direction has been uniquely derived by measurement. But is this really the case? In this situation the experiment outcome consists of the physical object 'screen' together with a physical object 'dot' on it. If we now come closer to the screen we may observe a chemical reaction, blurred across an area on the screen,  which gives rise to the visible dot. The dot, being a cloud of chemically interacting parts (these parts can be groups of molecules, for example), has many physical degrees of freedom and these degrees of freedom must be correlated with the measured particle because they {\it materialize} the experimental result. Recall that according to a widely accepted interpretation of quantum mechanics it is only through the experiment result that a quantum entity becomes a {\it real} physical object with a measured physical attribute. Hence, it does not make sense to say that the observed particle has only one degree of freedom ($+\onehalf$ or $-\onehalf$ in $z$-direction) because what we actually observe as a measurement result (by means of the chemically interacting cloud on the screen) is a physical system that has many more. Parts can therefore be viewed as the material constituents  of the physical object 'particle with spin $+\onehalf$ in $z$-direction'. Now many of these parts can be separated but at the same time some become indiscernible in our visual field no matter how close we observe the cloud because every time we {\it zoom-in} a new family of parts may emerge.  In this sense every experiment result can be partitioned such that the imperfection principle holds. We see that the principle is intimately related to our senses (not only to our visual sense but to all our senses that interact with the outer world) in that every physical experiment ultimately is an experiment carried out through our sensory apparatus; concurrently, our senses give rise to a collection of non separable outcomes of experiments. We return to this issue in section \ref{percep} where we will see that-- from a mathematically point of view-- any family of parts forms a complete ortholattice realized through a non-transitive binary relation called the proximity relation $P$. From a physical point of view we will recognize parts as a model of Poincar\'e's {\it physical continuum}.\medskip  

 Our second principle, called the {\it simulation principle}, says that the universe is a quantum automaton with a certain simulation running on it. The principle goes further in that it postulates that the aforementioned simulation executes a certain evolutionary process known from set theory. The set theory in question is Baltag's Structural Theory of Sets (STS) \citep{bal1999} which is a non-wellfounded set theory based on infinitary modal logic. Non-wellfounded set theories are logically weaker than conventional set theories (such as the classical Zermelo-Frankel-Axiom of Choice (ZFC) set theory): they do not have an Axiom of Foundation. Baltag's Structural Set Theory operates concurrently on two sides. On one side it operates with modal logic, i.e. a non-classical logic that allows for modalities of propositions such as possibility and necessity, while on the other side it represents all those sets that satisfy modal sentences by the so-called satisfaction axioms.  In STS, modal sentences are regarded as {\it analytical experiments} which means that each modal sentence is a possible statement about sets, and where the answer is the set that satisfies this statement. In this manner STS is an {\it analytical} set theory where sets are discovered as opposed to {\it synthetic} set theories,  such as ZFC, where sets are built recursively by means of the usual iterative concept of set. There is a natural process of {\it logical evolution} in STS that comes with a recursive formulation of modal sentences. The process is represented as a sequence of modal sentences ordered by instances of the ordinals. Each ordinal $\alpha$ gives rise to a stage of {\it structural unfolding} of a set. Thus {\it a priori} an arbitrary  set is completely unknown; instead it reveals its structure only step-wise through the successive stages of unfolding. The higher the stage ordinal $\alpha$ the better is our information about the analyzed set.\smallskip 

The simulation principle now postulates that the quantum automaton runs a simulation of the structural unfolding process of an initially unknown class $U$ called the universe. Structural unfolding generates a $\alpha$-sequence $\{\kri_U^\alpha\}$ of so-called Kripke models (or, Kripke structures) of modal logic. Loosely speaking, each Kripke model $\kri_U^\alpha$ is a rooted  graph with a finite number of labeled nodes and directed edges. The edges thus define a binary relation $R^\alpha_U$ on the set of nodes; it is referred to as the {\it accessibility relation} while the corresponding nodes are often termed {\it possible worlds}. In the case for $\kri_U^\alpha$ the accessibility relation is understood as the membership structure $\in$ between sets and their members (which in general are sets again)--thus every $\kri_U^\alpha$ canonically represents a tree. According to the simulation principle, it is this tree that is supposed to be simulated on the universal quantum automaton. For a more rigorous treatment of this situation we will show that this kind of simulation can be naturally represented by the well established mathematical concept of {\it bisimulation}. A bisimulation $\equiv_B$ is an equivalence relation between Kripke models. Indeed, bisimilar Kripke structures have the same modal language in that they share the same collection of modal sentences.  In order to establish the so-called {\it bisimulation principle}, we employ the previous imperfection principle and postulate the existence of a secondary Kripke model $\krip_\Sigma^\alpha$ in which the accessibility relation for Kripke models is given by a proximity relation $P_\Sigma^\alpha$. This is the first main step in our approach. It explicitly accounts for the inability to perform perfect physical experiments and it places this essential into the Kripke model $\krip_\Sigma^\alpha$. We then show that $\krip_\Sigma^\alpha$ becomes the quantum universe if it is concurrently identified as the immediate simulation of $\kri_U^\alpha$; and the attribute of being a ``immediate simulation'' we express mathematically through the concept of bisimulation. Thus the second step is to realize that both Kripke models, $\kri_U^\alpha$ and $\krip_\Sigma^\alpha$, are equivalent in the sense of bisimulation {\it although} both refer to different structures originally: $\kri_U^\alpha$ refers to an abstract membership structure in set theory while $\krip_\Sigma^\alpha$ refers to the structure of indiscernible outcomes of physical experiments. In this manner the preliminary simulation principle turns into its mathematical  form. It says: for all ordinal stages $\alpha$ of structural unfolding of the universe $U$ the Kripke models $\kri_U^\alpha$ and $\krip_\Sigma^\alpha$ are bisimilar, {\it viz.} $\kri_U^\alpha \equiv_B \krip^\alpha_\Sigma \,.$\bigskip   

Having introduced the basic principles we may now outline our strategy concerning the analysis the five main problems related to the quantum universe, i.e. questions (1.) -- (5.). First of all, in sections \ref{modal}, \ref{dempster}, and in \ref{baltag}, we give an eclectic summary of the mathematical methods we need to employ. This summary involves modal logic, Dempster-Shafer theory and its relation to modal logic, and Baltag's structural theory of sets. In section \ref{percep} we introduce the so-called quantum sets, which is another term  for parts of perception as mentioned earlier in this introduction. We explain how parts are related to imperfect experiments. Following the works of J. L. Bell \cite{bel1986, bel2000}, we show how quantum sets are used canonically to model quantum logic.\medskip

The main part the present text consists of the sections \ref{preferred bases} and \ref{metric spaces}. Both parts contain quite different mathematical aspects and therefore we put them in two different sections. However, both sections belong together {\it logically} because they constitute an inductive proof of our assertion according to at any ordinal stage of the quantum universe there is a preferred basis.

In doing the inductive proof, we first take for granted the existence of an appropriate Hilbert space structure $\hil^\alpha$ along with an orthonormal basis $b^\alpha$ of eigenvectors of a Hermitian homeomorphism on $\hil^\alpha$; this happens in sections \ref{hilbert} and  \ref{bases and states}. The induction step, i.e. the proof that this is also the case for the successor stage $\alpha+1$, is made throughout the rest of the present article. The proof step begins in section \ref{unfoldphys} where the Kripke structure $\kri_U^\alpha$ is introduced; in this section we also discuss the role of the structural unfolding process as the main evolutionary process of the universe. In section \ref{preferred choice}, then, we derive a precise formulation of the Bisimulation Principle. As a direct application of the Bisimulation Principle we quote our main result, Proposition \ref{bpresults}, which finishes the inductive proof. 

In section \ref{metric spaces} we study the discrete tree metric structure that follows naturally from the Bisimulation Principle. We find that tree metrics can be isometrically embedded into the normed spaces $L_1$ and $l_1$. In this context we stress certain similarities to error correcting codes.  This characterization allows for a non-necessarily isometric embedding of the tree metric into the Euclidean space $l_2$. The construction of an Euclidean distance matrix out of the embedded vectors in the Hilbert space $l_2$ is the last step; the distance matrix represents the self-test $\Sigma^\alpha$ and its eigenvectors form a non-degenerate basis of $\hil^\alpha$. This step completes our inductive proof in that it justifies Proposition \ref{bpresults} {\it a posteriori}.\medskip 

In section \ref{conclusions}, we close with several comments and remarks about our results.  

\section{Basics of modal logic \label{modal}}
Modal logic is a natural extension of classical, i.e. binary, logic. It is concerned with logical facts, such as logical inferences, that involve modalities, i.e. qualifications of propositions. Its language consists of atomic propositions $\varphi, \psi, \ldots$, of logical connectives $\neg, \wedge, \vee, \Rightarrow, \Leftrightarrow$, of modal operators of {\it necessity} $\Box$ and {\it possibility} $\Diamond$, and of supporting symbols $(,), \, \{,\}$. The main objects of interest are {\it formulas} or {\it sentences}: any atomic proposition is a formula; and if $\varphi$ and $\psi$ are formulas, then so are $\neg \varphi,\,  \varphi \wedge \psi, \, \varphi \vee \psi,\, \varphi \Leftarrow \psi, \, \varphi \Leftrightarrow \psi, \, \Box \varphi, \, \Diamond \varphi$. 

Modal languages often are introduced as formal systems equip\-ped with extra axioms or with additional inference rules; however, the common basic structure of modal languages as introduced above will be sufficient for now. 
Each formula assesses a meaning through its {\it truth value} given in a context. These contexts are expressed in terms of {\it models} of modal logic. A model, $\mathbf{M}$, of modal logic is the triple
$$
{\mathbf M} = \langle W, \, R, \, V\rangle\,,
$$
where $W$ is a set of possible worlds, $R$ is binary relation on $W$, and where $V$ is a set of value assignment functions, one for each world in $W$, by which a truth value of truth $(T)$ or of falsity $(F)$ is assigned to each atomic proposition. The triple $\langle W, R, V \rangle$ is also referred to as {\it Kripke structure} (other frequently used terms are {\it Kripke model of modal logic} or {\it Kripke semantics for modal logic}). Value assignment functions are inductively applied to all constructible formulas in the usual way, the only interesting cases being
\begin{equation}
\label{nec}
v_i(\Box\varphi) = T \quad \mbox{iff} \quad \forall\, w_j \in W: \,w_iR w_j\quad\mbox{implies}\quad v_j(\varphi) = T \,, 
\end{equation}
with $v_i, v_j \in V$ and $w_i, w_j \in W$ for all indices $i, j$, and
\begin{equation}
\label{pos}
v_i(\Diamond\varphi) = T \quad\mbox{iff}\quad \exists \, w_j \in W: \,w_iR w_j\,\,\,\mbox{implies}\,\,\, v_j(\varphi) = T \,.
\end{equation}
The binary relation $R$ is called {\it accessibility relation}; we say that the world $w_j$ is accessible to world $w_i$ when $w_iR w_j$. We assume that $W$ is finite and that its cardinality is denoted by a natural number $N = |W|$. It is convenient to denote $ W = \{w_1, w_2, \ldots, w_N\}$ and to represent the relation $R$ by an $N \times N$ matrix ${\mathbf{R}} = [r_{ij}]$, with components
$$
r_{ij} = \left\{ \begin{array}{rcl}
1 & \mbox{if} & w_iR w_j\,,\\
0 & \mbox{if} & \neg (w_iR w_j) \,.
\end{array}\right.
$$
Additionally, for a formula $\varphi$, we will often write $v_i(\varphi) = 1$ when $v_i(\varphi) = T$, and $v_i(\varphi) = 0$ when $v_i(\varphi) = F$. For further purposes, we consider also a weighting function $\Omega$. This function becomes a component of the model $\mathbf{M}$ and maps possible worlds into the real interval $[0,1]$ so that
$$
\sum_{i = 1}^N \Omega(w_i) = 1 \,.
$$
It is also useful to denote $\Omega(w_j)$ as $\omega_j$.

\section{Dempster-Shafer theory and modal logic \label{dempster}}

Dempster-Shafer theory (often referred to as Evidence Theory) can be titled a mathematical description of  {\it belief} \citep{sha1976}. The latter is a generalization of the mathematical concept of probability. An intuitive approach is first to look on probabilities in their elementarity, and then to mathematically generalize the concept of the notion of {\it point probability} towards {\it set probability}. Whenever a condition $H \in Y$  is at place, it may give rise to a probability for an event $e \in X$ to happen; this situation normally is expressed by a mapping $P(e|H)$, where $X$ and $Y$ are assumed to be finite sets. Also, $P$ obeys Kolmogorov's axioms of probability and returns a probability value for any of the elements $x \in X$ which altogether are interpreted as mutually exclusive events. Thus the domain of $P$ consists of individual elements or simply (labeled) points of the set $X$. With the additivity law for probabilities at hand, $P$ is used to generate a probability measure on the power set of $X$, ${\mathcal P}(X)$. Dempster-Shafer theory now takes on a broader view and considers already from the beginning a real valued mapping $m$ with domain ${\mathcal P}(X)$. This mapping is the {\it basic probability assignment} sharing the following properties
\begin{eqnarray}
\nonumber
m(\emptyset) &=& 0 \,,\\
\sum_{A \in {\mathcal P}(X)} \!\!m(A) &=& 1\,.
\end{eqnarray}
A set $A \in  {\mathcal P}(X)$ with $m(A) > 0$ is called a {\it focal element}. Given a basic probability assignment $m$ and a set $A \in  {\mathcal P}(X)$ one defines the total belief of $A$ as
$$
\bel(A) := \sum_{B \subseteq A} \!m(B) \,. 
$$
$\bel$ is a measure on ${\mathcal P}(X)$ and it is called the belief measure. The dual value of $\bel(A)$ is given through its total {\it plausibility} 
$$
{\rm Pl}(A) := 1 - \bel(c(A))\,, 
$$
where $c$ denotes the set complement of $A$. Likewise, ${\rm Pl}$ is the plausibility measure. Although Bel turns out to be a measure ${\mathcal P}(X)$ it does not, in general, follow the rules of probability measures. One main difference is that Bel generally obeys the {\it super-additivity rule} rather than the ordinary additivity law for probability measures (In its dual analogy, Pl has the {\it sub-additivity property}.). Thus given two sets $A, B \subseteq X$  we have
$$
\bel(A \cup B) \geq \bel(A) + \bel(B) - \bel(A \cap B)\,.
$$ 
In \citep{rkp1999} it is argued that since in quantum mechanics we naturally come into situations where ``probability measures'' do no share the additivity rule,\footnote{Think of a probability $P_{12}$ of a state $\psi_{12}$ that is a non-trivial  superposition of  two quantum states $\psi_1$ and $\psi_2$, then $P_{12} \neq |\psi_1|^2 + |\psi_2|^2$. On the other hand, when we wish to obtain the total probability $P_{12}$ of those two states not being in a superposition we have $ P_{12} =  |\psi_1|^2 + |\psi_2|^2$.} belief and plausibility measures of Dempster-Shafer theory describe ``probabilities'' in quantum mechanics more naturally. We follow this idea and show in section \ref{preferred bases} of this paper that Dempster-Shafer theory of evidence indeed becomes the proper conceptual framework when dealing with amplitudes of the $\psi$-function.\bigskip

We close this section by quoting a result due to \citep{rkhc1996}. This result establishes a relation between Dempster-Shafer Theory and semantics of modal logic. In fact, Dempster-Shafer theory can be represented in terms of a model of modal logic if we employ propositions of the form\medskip

$\varphi_{_A}$ : ``A given incompletely characterized element $\epsilon \in X$ is characterized as an element $\epsilon \in A$''\,,\\[.2 cm]
where $A \in {\mathcal P}(X)$. It is then sufficient to consider as atomic propositions only propositions $\varphi_{\{x\}}$, where $x \in X$. General propositions of the kind $\varphi_A$ are then defined by the formulas
$$
\varphi_{_A} = \bigvee_{x \in A} \varphi_{_{\{x\}}}
$$
for all $A \neq \emptyset$ and
$$
\varphi_{_\emptyset} = \bigwedge_{x \in X} \neg \varphi_{_{\{x\}}} \,.
$$

For each world, $w_i \in W$, of a Kripke structure it is assumed that $v_i(\varphi_{{\{x\}}}) = 1$ holds for one and only one $x \in X$; this property is called singleton valuation assignment (SVA) \citep{vtb1999}. Also, the relation $R$ is assumed to be reflexive or at least {\it serial}, i.e. for all $w_i \in W$ there is a $w_k \in W$ such that $w_iRw_k$. Under this assumption Resconi {\it et al.} \citep{rkhc1996} propose a modal logic interpretation of the basic functions in Dempster-Shafer-Theory:
\begin{eqnarray}
\bel(A) &=& \sum_{i = 1}^N \omega_i \, v_i(\Box \varphi_{_A})\,,\\
\pl(A) &=& \sum_{i=1}^N \omega_i \, v_i (\Diamond \varphi_{_A})\,,\\
\label{bpa}
m(A) &=& \sum_{i=1}^N \omega_i \, v_i\!\!\left(\Box\varphi_{_A} \wedge \left[ \bigwedge_{x \in A} \Diamond \varphi_{_{\{x\}}}\right ]\right )\,.
\end{eqnarray}
In section \ref{borns rule} we will demonstrate how $m(A)$ in equation \ref{bpa} is used to establish a modal interpretation of Born's rule.

\section{Baltag's structural theory of sets \label{baltag}}

In his seminal work \citep{bal1999} Baltag constructs a non-wellfounded, universal set theory based on a {\it structural} conception of sets. Briefly stated, a non-wellfounded set theory is a set theory where the membership relation $\in$ is not wellfounded as opposed to wellfounded set theories like Zermelo-Fraenkel (ZF) set theories which include the Axiom of Foundation. Several non-wellfounded set theories have been proposed by means of additional existence axioms since in 1954 Bernays proved the relative independence of the Axiom of Foundation in ZF \citep{ber1954}. Systematic constructions of non-wellfounded set theories by introducing so-called {\it Antifoundation Axioms} (AFA) date as far back as 1926 when Finsler introduced the {\it Finsler}-AFA (FAFA) in set theory \citep{fin1926}. In non-wellfounded theories ``exotic'' sets like
$$
a = \{b, a\} \quad\mbox{or}\quad a = \ldots\{\{\{b \}\}\}\ldots
$$
may appear. Such sets are often called {\it hypersets} and are used to represent {\it self-referential} structures or situations because a non-wellfounded set may---for example--well become a member of its own member. A structural understanding of sets is dual to the classical iterative (i.e., synthetic) concept of set. While in the latter we consider sets as built from some previously given objects in successive stages, the former presupposes that {\it a priori} a set is a {\it unified totality} that reveals its abstract membership structure only step by step through the process of structural unfolding. This stepwise discovery of the set structure is generated by imposing questions (which Baltag calls analytical experiments) to the initial object; the answers to these questions are the stages of structural unfolding.  The idea behind it is that sets are what is left when we take an aggregate (a complex object, to say) and we abstract everything but its membership structure. This structure is {\it pointed}, in that it has a root: the underlying process of unfolding the structure, by successive decompositions, has a starting point, namely the very object under consideration. Thus sets are here conceived as {\it pointed binary structures}; this is the same as considering a Kripke structure with a distinguished set (i.e., the root) and having an accessibility relation representing membership. Thus loosely speaking, many sets may be conceived simply as pointed, directed graphs. At a given stage of unfolding, which is labeled by an ordinal $\alpha$, we have only {\it a partial} description of the set considered. Let this arbitrary set be $a$, and let $a^\alpha$ be the present stage of unfolding; then, in order to obtain the next stage of unfolding, we take the set of all $\alpha$ unfoldings of the members of $a$. For limit stages $\lambda$ (i.e., when $\lambda$ is a limit ordinal), suppose we are given all $\alpha$ unfoldings $a^\alpha$ of $a$ with $\alpha < \lambda$ (Although we are going to quote the unfolding rule for limit stages $\lambda$, all methods presented in this work will refer to finite ordinals only.)  Observe that there is already a temporal metaphor within: there is a ``logical'' concept of time, given by a succession of stages of structural unfolding. Now naively, the unfolding process can be defined by the following recursion on the ordinals: for every ordinal $\alpha$ and every set $a$, the {\it unfolding of rank $\alpha$} is the set $a^\alpha$, given by
\begin{eqnarray}
\nonumber
a^{\alpha +1} &=& \{b^\alpha : b \in a\}\\
\nonumber
a^\lambda &=& \langle a^\alpha \rangle_{\alpha < \lambda} \, \mbox{, for limit ordinals,} \, \lambda\,.
\end{eqnarray} 
Surely, this definition is meaningful for all wellfounded sets, but for a larger objects it is inappropriate in general. Larger objects are general {\it pointed systems}, i.e. a generalization of the concept of graphs in which the collections of all pairs of nodes may form proper classes. Since $\in$-recursion is equivalent to the Axiom of Foundation, $\in$-recursion as introduced above is in general not appropriate for pointed systems that are proper classes also.

To find a definition of structural unfolding for more general objects, i.e. all pointed systems or proper classes,  Baltag takes seriously the fact that at every ordinal stage we can only have a partial description of a system. This description is realized through formulas in modal language describing the membership structure at a certain stage of unfolding. An essential ingredient here is the {\it observational equivalence between systems}. Observational equivalence is given by an equivalence relation on modal formulas; this notion does therefore not refer to the intended sets or classes directly, rather it studies the underlying language of modal sentences associated to these sets or classes. It turns out that with {\it infinitary modal logic} observational equivalence between arbitrary pointed systems can be naturally defined. In STS, a modal theory $th(a)$ for every set $a$ is constructed through the so-called satisfaction axioms. Before we quote these axioms we may first introduce the underlying modal language.
\begin{enumerate}
\item Negation. Given a possible description $\varphi$ and an object $a$, we construct a new description $\neg \varphi$, to capture the information that $\varphi$ does not describe $a$.
\item Conjunction. Given a set $\Phi$ of descriptions of the object $a$, we accumulate all descriptions in $\Phi$ by forming their conjunction $\bigwedge\Phi$.
\item Unfolding. Given a description $\varphi$ of some member (or members) of a set $a$, we {\it unfold} the set by constructing a description $\Diamond\varphi$, which captures the information that $a$ has some member described by $\varphi$.
\end{enumerate}
The language generated by these three rules and which allows for infinitary conjunctions is called {\it infinitary modal logic}, $L_\infty$. With $\bigvee$ and $\Box$ as the duals to $\bigwedge$ and $\Diamond$, respectively, we introduce some other operators:
\begin{eqnarray}
\nonumber
\Diamond \Phi &=:& \{\Diamond \varphi: \varphi \in \Phi\}\,,\\
\nonumber
\Box\Phi &=:& \{\Box\varphi : \varphi \in \Phi\}\,,\\
\nonumber
\varphi \wedge \psi &=:& \bigwedge\{\varphi, \psi\}\,,\\
\nonumber
\varphi\vee\psi &=:& \bigvee\{\varphi, \psi\}\,,\\
\nonumber
\bigtriangleup\Phi &=:& \bigwedge\Diamond\Phi \wedge \Box \bigvee\Phi\,.
\end{eqnarray}
The satisfaction axioms presume the existence of a class a $Sat$; each element of $Sat$ is a pair of a set $a$ and a modal sentence $\varphi$. Writing $a \models \varphi$ for $(a, \varphi) \in Sat$, these axioms read as
\begin{eqnarray}
\nonumber
&{\rm{(SA1)}}& \quad\quad a \models \neg \varphi \quad {\rm{iff}} \quad a \not\models \varphi\\
\nonumber
&{\rm{(SA2)}}& \quad\quad a \models \bigwedge \Phi \quad {\rm{iff}} \quad a \models \varphi\quad \mbox{{\rm for all}} \quad\varphi \in \Phi\\
\nonumber
&{\rm{(SA3)}}& \quad\quad a \models \Diamond\varphi \quad {\rm{iff}} \quad a' \models \varphi\quad \mbox{{\rm for some}}\quad a'\in a
\end{eqnarray}       
Thus given a set $a$ its theory, $th(a)$, consists of all modal formulas satisfied by $a$ by means of these axioms. With this setting the notion of unfolding of a set $a$ admits now an expression through modal sentences $\varphi^\alpha_a$ defined for any cardinal number $\alpha$ as
\begin{eqnarray}
\label{unfold}
\varphi^{\alpha+1}_a &=:& \bigtriangleup\{\varphi^\alpha_b : b \in a\}\,,\\
\varphi^\lambda_a &=:& \bigwedge\{\varphi_a^\beta:\beta<\alpha\}\quad {\rm for \,limit \, cardinals}, \lambda\,.
\end{eqnarray} 
Unfoldings of rank $\alpha$ are {\it maximal} from an informational point of view as they gather all the information that is available at stage $\alpha$ about a set and its members. In formal language this statement reads as the proposition: $b \models \varphi^\alpha_a$ iff $b^\alpha = a^\alpha$. Now this enables us to explain what we mean by {\it observationally equivalent}: 

{\it two sets, classes or systems are said to be observationally equivalent if they satisfy the same infinitary modal sentences, i.e. if they are modally equivalent.}

We close our short introduction with two general remarks stressing the beauty of the Structural Theory of Sets. First, a model $U$ of this set theory can be seen as the largest extension of a model $V \subset U$ of ordinary Zermelo-Fraenkel-Axiom of Choice (ZFC) set theory that still preserves the property of modal characterization. And second, STS belongs to {\it circular model theory}, in the sense that it contains its own model as an object in $U$.\smallskip 

Structural unfolding of an arbitrary set or class by means of modal descriptions is the key for our further investigations in this article. Indeed, we are going to employ the structural unfolding rule (\ref{unfold}) as a model of the evolutionary process of the quantum universe.

\section{Perception, experiments,  and quantum logic \label{percep}}

The motivating question for this section is whether continuity of perceived experimental outcomes generally implies that the observed physical matter has to be continuous, too. This question and the negative opinion about it, saying that this implication is by no means necessary, has its own history. Poincar\'e, for instance, made a clear distinction between the {\it physical continuum} and the {\it mathematical continuum}\footnote{The author became recently aware of the work of M. Planat \citep{pla2004} where an interpretation of the perception of time is given on the ground of Poincar\'e's ideas.}-- he writes in 1905 \citep{poi1905}:

\begin{quote} 
{\it
We are next led to ask if the idea of the mathematical
continuum is not simply drawn from experiment. If that be so, the rough data of
experiment, which are our sensations, could be measured [...] It has, for instance, been observed that a weight A of 10 grammes and a weight B of 11 grammes produced identical sensations, that the weight B could no longer be distinguished from a weight C of 12 grammes, but hat the weight A was readily distinguished from the weight C. Thus the rough results of the experiments may be expressed by the following relations: $A = B , \,B = C , \, A < C$, which may be regarded as the formula of the physical continuum. But here is an intolerable disagreement with the law of contradiction, and the necessity of banishing this disagreement has compelled us to invent the mathematical continuum. We are therefore forced to conclude that this notion has been created entirely by the mind, but it is experiment that has provided the opportunity. We cannot believe that two quantities
which are equal to a third are not equal to one another, and we are thus led to suppose that A is different from B, and B from C, and that if we have not been aware of this, it is due to the imperfections of our senses. [...]\\
What happens now if we have recourse to some instrument to make up for the weakness of our senses? If, for example, we use a microscope? Such terms as A and B, which before were indistinguishable from one another, appear now to be distinct: but between A and B, which are distinct, is intercalated another new term D, which we can distinguish neither from A nor from B. Although we may use the most delicate methods, the rough results of our experiments will always present the characters of the physical continuum with the contradiction which is inherent in it. We only escape from it by incessantly intercalating new terms between the terms already distinguished, and this operation must be pursued indefinitely. We might conceive that it would be possible to stop if we could imagine an instrument powerful enough to decompose the physical continuum into discrete elements, just as the telescope resolves the Milky Way into stars. But this we cannot imagine; it is always with our senses that we use our instruments; it is with the eye that we observe the image magnified by the microscope, and this image must therefore always retain the characters of visual sensation, and therefore those of the physical continuum.
}
\end{quote}
\medskip

We want to explore the possibility of a physical continuum on a formal level. Let $X$ denote a set of finite cardinality $N$ representing  mutually exclusive events that in our context represent the set of all possible outcomes of a physical experiment, and let ${\mathcal P}(X)$ be its power set. A {\it proximity relation} $P$ is a binary relation between the elements of $X$ that is reflexive and symmetric, but not necessarily transitive \citep{bel2000, bel1986}. We call the pair $(X, P)$ the proximity space. For each $x \in X$ the set
$$
Q_x = \{y \in X : \,xPy\}
$$
is called a {\it quantum associated to $x \in X$}. Then the set of all quanta, $ {\mathcal R} = \{ Q_x | x \in X\}$, is called the reference set of the proximity space considered. Within the reference set, quanta are the smallest recognizable subsets of $X$. Any subset of $X$ that is a union of of some quanta is called a {\it quantum set} (or, a {\it part} in Bell's terminology). We denote the set of all quantum sets as ${\mathcal Q}_P$.

Within our interpretation of $(X,P)$, the proximity relation characterizes the {\it indistinguishableness} of outcomes due to experimental errors. Experimental errors too are present in our ability to perceive nature through our sensory fields generated by vision, touch, sound, and smell. Experimental errors in this sense are an inherent feature of our limited ability to receive information from nature. They manifest the impossibility to prepare and to perform an experiment providing us with outcomes of unlimited precision. This natural limitation, which is due to an imperfect knowledge about the experiment setup and due to an limited control of the experiment process, is expressed through quantum sets. In this context, we say that a physical experiment, or a perceptual process, with possible outcomes in $X$ is {\it imperfect} if there is $x, y \in X$, with $x \neq y$, such that $xPy$.\medskip  

According to Bell's work, quantum sets can be used immediately to construct a model of quantum logic. In fact, the set ${\mathcal Q}_P$ can readily be interpreted as  a complete ortholattice, that is a tuple $\orthop = ({\mathcal Q}_P, \cap_P, \cup_P, ^\perp)$, if we equip  ${\mathcal Q}_P$ with a join operation $\cup_P$ taken as the usual set-theoretic union, with a meet operation $\cap_P$ of two quantum sets as the union of all quanta in their set-theoretical intersection, and with an unary relation $^\perp$ with
$$
^\perp Q = \{y \in X | \, (\exists x \notin Q)(xPy)\}
$$
for any $Q \in {\mathcal Q}_P$. A complete ortholattice is known to be a proper model for quantum logic in the sense of von Neumann and Birkhoff \cite{bn1963}. However, it has not been introduced here as a lattice of closed subspaces of a Hilbert space but rather as a lattice of quantum sets (or, parts) for a given proximity space $(X, P)$. In this manner proximity relations can be viewed as an alternative entry to the quantum realm-- as has been proposed by Bell \citep{bel1986}. One may now directly recover observables, for instance, as the complete ortholattice $\orthop$ naturally extends to a  {\it proposition system}; physical observables are then defined through $c$-morphisms from a complete Boolean algebra into the proposition system $\orthop$ (for details, see \citep{pir1976} or, more recently, \citep{sk2003}).\smallskip 

It is evident that not all sets in ${\mathcal P}(X)$ are quantum sets; nevertheless a given proximity relation offers a mathematical classification of any two sets in ${\mathcal P}(X)$.  Given $A, B \in {\mathcal P}(X)$,  we say $A$ and $B$ are {\it separated} if $A \cap B = \emptyset$ and if for all $x \in A$ it is $Q_x \cap B = \emptyset$; and due to the symmetry of the proximity relation the same holds for the elements of $B$. Generally, for any two sets $A$ and $B$ which are not separated one distinguishes two cases: superposition and incompatibility. For an introduction of these cases we refer to the work of Resconi {\it et al.} \citep{rkp1999}, or, on a more fundamental level,  to Bell's original work \citep{bel1986, bel2000}.\footnote{In \citep{bel2000} Bell uses this classification to demonstrate that the human visual field resembles quantum behavior in terms of superposition.} Both classes resemble situations in Hilbert spaces of quantum systems where two states may arise in a linear superposition, and where two observables may be incompatible. We stress that separability, superposition and incompatibility on a complete ortholattice generated from a proximity space is in general not the same as separability, superposition and incompatibility arising on a complete ortholattice associated to closed subspaces of the quantum system's Hilbert space. There may arise several different ortholattices for one given quantum system. For example, let $\hil$ be a separable Hilbert space, then we have the complete ortholattice ${\mathcal L_\hil}$ of closed subspaces in $\hil$ but at the same time we may obtain another complete ortholattice $\orthop$ as follows. Let $\Sigma$ be a Hermitian operator on $\hil$ admitting an orthonormal basis $b \subset \hil$ of non-degenerate eigenvectors of $\hil$. Each eigenvector  $x \in b$ corresponds to a measurement outcome documented with the associated eigenvalue $\lambda_x \in {\mathbb R}$. We define the proximity relation $P$ for all $x, y \in b$
$$
(xPy) \quad\mbox{iff}\quad \mbox{($\lambda_x$ is indistinguishable from $\lambda_y$ by experiment)}.
$$
As mentioned earlier in this section, such a proximity relation can be readily used to define $\orthop$. And, clearly, there is no necessity to imply that ${\mathcal L_\hil}$ and $\orthop$ are isomorphic ortholattices in any plausible sense. Moreover, the present approach to quantum logic via proximity spaces turns out to be a general method. This follows from the fact that {\it all} complete ortholattices $\mathcal L_\hil$ representing closed subspaces of a separable Hilbert space $\hil$ are isomorphic (as ortholattices) to proximity spaces based on the proximity relation
$$
(sPt) \quad\mbox{iff}\quad (s,t) \neq 0\,,
$$
for all $s,t \in \hil \backslash \{0\}$, and where $(.,.)$ is the inner product on $\hil$ \citep{bel1986}. It is in  this sense that proximity relations give a general approach to the mathematical foundations of quantum physics.\medskip

We may now incorporate modal logic and Kripke semantics into this framework. This can be done first by identifying the basis elements $x \in b$ as the {\it possible worlds}, and the proximity relation $P$ as the {\it accessibility relation}g between the latter.  Value assignment functions are then constructed as follows: for any $x \in b$ we set  $v(\varphi_{Q_x})=T$, and, moreover, for an arbitrary subset $ a \subseteq b$ we have $v(\varphi_a)=T$ if $Q_x \subseteq a$, and $v(\varphi_{Q_x})=F$ if $Q_x \nsubseteq a$. Here, again, $\varphi_a$ denotes the proposition ``$\varphi_a$: An element $x \in b$ is characterized as $x \in a$'' (see, section \ref{dempster}). Hence, we have established a Kripke structure on the elements of a given Hilbert space basis $b$ by identifying the proximity relation $P$ with the accessibilty relation $R$ in Kripke semantics.\medskip

A physically realized proximity space $(X, P)$ thus enables us to construct a semantic interpretation of quantum logic. It is a general and  alternative approach to the quantum world as it provides us quite naturally with fundamental quantum concepts such as complete ortholattice, separation, superposition, and incompatibility. It can be used also to model the indistinguishableness of experimental results such as parts in the perceptual field without making explicit use of the mathematical continuum ${\mathcal P}(\aleph_0)$. From this point of view a proximity relation can be regarded as a mathematical formulation of Poincar\'e's physical continuum. And, when applied to set of possible outcomes of a quantum measurement, the proximity relation determines a Kripke structure where the possible worlds coincide with possible measurement outcomes.

\section{Preferred bases and state selection\\ in the quantum universe \label{preferred bases}}

As described in the introduction, we consider the universe as a large quantum system that proceeds step-wise from one stage to the successive stage. Our aim is to reconcile our methods described in order to show by induction over the stage ordinals $\alpha$ that a preferred basis emerges at {\it all} stages of the quantum universe. To achieve this goal, we first formulate the proof statement for $\alpha$ in the sections \ref{hilbert}, \ref{bases and states}, and \ref{modal satis}. In the latter section we also show the validity of the base case $\alpha=0$. The induction step begins with section \ref{unfoldphys} and it is completed with Proposition \ref{bpresults} in section \ref{preferred choice}. Sections \ref{borns rule} and \ref{state selection} contain further implications regarding Born's rule and state selection.

Each of the following sections begins with a main part that ends with a ``$\square$'' sign; further comments and explanations are placed thereafter.\medskip

\subsection{Hilbert space \label{hilbert}}
At any ordinal stage $\alpha$, there is a complex Hilbert space $\hil^\alpha$ with a number of dimensions $\dim \hil^\alpha = N_\alpha$.\boxf\medskip

This assumption is a recapitulation of our original working hypothesis in that the quantum universe is represented as a large but finite quantum register realized as a complex Hilbert space of finite dimension.\footnote{We will often call the whole Hilbert $\hil^\alpha$ a quantum register or, simply, register.}\medskip

\subsection{Bases and states \label{bases and states}}
At each stage $\alpha$, there is a distinguished pure state $\Psi_\alpha \in \hil^\alpha$ called the state of the universe at stage $\alpha$. Additionally, $\Psi_\alpha$ is an element of a basis $b^\alpha \in \bases$, where $\bases$ is the set of bases of $\hil^\alpha$. Moreover,  $b^\alpha$ forms the family of non degenerate eigenvectors of a Hermitian Hilbert space homeomorphism $\Sigma^\alpha: \hil^\alpha \rightarrow \hil^\alpha$ called the self-test of the universe at stage $\alpha$.\boxf\medskip

We remark that with the above assumptions stage dynamics, when seen as the procedure of taking the step from one stage represented by $\Psi_\alpha$ to the next stage represented by $\Psi_{\alpha+1}$, requires a process capable of selecting a unique element from the preferred basis $b^{\alpha+1}$. But such a process intended cannot be accomplished through usual {\it no-collapse} approaches to the measurement problem-- such as through the decoherence method or through the many-worlds approach-- as those are unable to explain the selection of an individual state from the family of pointer states, i.e. the apparent collapse of the wave-function. Empirically, a collapse of the wave-function is  commonly experienced by human subjects conducting a measurement. Moreover, as has been objected for the many-worlds approach \citep{sta2002}, these methods even may fail to depict a preferred basis beforehand. Thus not only it becomes unclear how to select a single state, but it is also not evident how the set of alternatives looks like where any (hypothetical) selection process is about to act on. In contrast to this situation, we propose an approach in which the emergence of a preferred basis {\it and} the selection of a distinguished state out of the elements of the preferred basis occur naturally at every stage. This occurrence may be called a {\it self-organization of the measurement}. The crucial question therefore is how this particular self-organization of the measurement process is ever achieved within the realm of the quantum universe. Are there, for instance, rules or principles that control the intended emergence?



\subsection{Modal satisfaction \label{modal satis}}
At each stage $\alpha$ there is a modal sentence $\varphi_U^\alpha$ such that it is satisfied by the $\alpha$-unfolding $U^\alpha$ of $U$, i.e. $U^\alpha \models \varphi_U^\alpha$. Further, there is a function $Z_U$ from the finite ordinals to the natural numbers such that
\begin{enumerate}
\item if $\beta < \alpha$ then $Z_U(\beta) \leq Z_U(\alpha)$, and
\item $Z_U(\alpha) = \dim \hil^\alpha$ for all finite ordinals $\alpha$.
\hfill\boxf\medskip 
\end{enumerate}

This completes our collection of assumptions for the inductive proof. What remains to be done is to show their validity in the $\alpha+1$ case. We note that the base case for the inductive proof, $\alpha = 0$, is verified readily with $\hil^0 = \{c\,1: c \in {\mathbb C}\}$, with $(a, b) : = \bar{a}\,b$ for all $a,b \in \hil^0$, with the preferred basis $b^0 = \{1\}$, and with the Hermitian operator $\Sigma^0 = {\rm id}_{\hil^0}$.\smallskip

\subsection{Structural unfolding rule \label{unfoldphys}}

The universe as a whole is an unknown class $U$ and it reveals its structure only partially through a stage-wise unfolding process. At a given ordinal stage $\alpha$, the universe $U$ unfolds due to the rule
\begin{eqnarray}
\label{unfoldrule}
\varphi^{\alpha+1}_U &=:& \bigtriangleup\{\varphi^\alpha_u : u \in U\}\\ 
\nonumber
&=& \bigtriangleup D^{\alpha+1}\,,
\end{eqnarray}
with $D^{\alpha+1} = \{\varphi^\alpha_u : u \in U\}$. This is the structural unfolding rule of STS, c.f. equation (\ref{unfoldrule}). And this rule enables us to construct a Kripke structure $\kri_U^{\alpha+1} = \langle W_U^{\alpha+1}, R_U^{\alpha+1}, V_U^{\alpha+1}\rangle$ with a distinguished world $w^* \in W^{\alpha+1}_U$, i.e. with a $w^* \in W^{\alpha+1}_U$ having the property
$$
(\nexists w \in W^{\alpha+1}_U)\,(wR_U^{\alpha+1} w^* \,\,\mbox{and}\,\, w \neq w^*)\,,
$$
which we call the {\it point property} of $w^*$. This construction goes as follows. Given an unfolding $U^{\alpha+1} \models \varphi^{\alpha+1}_U$, consider a series of {\it children sets} defined as 
\begin{eqnarray}
\nonumber
\ch^1(U^{\alpha+1}) &:=& \{\varphi^\alpha_{u_1}: {u_1} \in U^{\alpha+1}\}\,,\\
\nonumber
\ch^2(U^{\alpha+1}) &:=& \{\varphi^{\alpha-1}_{u_2}: u_2 \in u_1 \,\,\mbox{and}\,\,u_1 \in U^{\alpha+1}\}\,,\\
\nonumber
&\vdots&\\
\nonumber
\ch^{\alpha+1}(U^{\alpha+1}) &:=& \{\varphi^{0}_{u_{\alpha+1}}: u_{\alpha+1} \in u_{\alpha} \,\,\mbox{and}\,\ldots\,\,\mbox{and}\,\,u_1 \in U^{\alpha+1}\}\,.
\end{eqnarray}
Then let us define the function
\begin{equation}
\label{nst}
Z_U(\alpha+1) := 1 + \sum_{i=1}^{\alpha+1} |\ch^{i}(U^{\alpha+1})| = N_{\alpha+1} = N'\,.
\end{equation} 
It follows that $Z_U$ is a monotonic function, because the $\alpha+1$ unfolding is satisfied by the set $U^{\alpha+1}$ whose members are all sets available that satisfy the modal formula of the $\alpha$th unfolding. In this  way the first condition in section (\ref{modal satis}) is already met. Equation (\ref{nst}) allows us to label uniquely each modal formula $\varphi_n^i$ with a number $n \in \{1,\ldots,N'-1\}$ and with a determined {\it stage number} $i \in \{0,\ldots,\alpha\}$ (Observe that in our construction a choice of $n$ already determines the value of $i$.). Additionally, we choose the only remaining formula $\varphi^{\alpha+1}_U$ to be associated with the number $N'$; in this way the distinguished world $w^*$ corresponds to $\varphi^\alpha_{N'} = \varphi_U^{\alpha+1}$, i.e. we set ${\rm ch}^0(U^{\alpha+1}) = \{\varphi^{\alpha+1}_{N'}\} = \{\varphi_U^{\alpha+1}\}$. So we have $|W_U^{\alpha+1}| = N'$ and each world $w \in W_U^{\alpha+1}$ uniquely corresponds to a formula $\varphi^i_n$ and the collection of all such formulas constitutes the set of possible worlds $W_U^{\alpha+1}$  with
$$
W_U^{\alpha+1} = \bigcup_{i=0}^{\alpha+1} {\rm ch}^i(U^{\alpha+1})\,.
$$

 A pointed graph $(W_U^{\alpha+1}, R_U^{\alpha+1})$ is constructed by means of the binary relation
$$
(w_nR_U^{\alpha+1} w_m) \,\,\mbox{iff}\,\,\, (\exists i \in \{1,\ldots,\alpha+1\}) \,(\varphi^i_n =: \bigtriangleup D^i \,\,\mbox{and}\,\, \varphi^{i-1}_m \in D^{i}),
$$
where $n, m \in \{1,\ldots,N'\}$. Further, to each $n_a \in \{1,\ldots,N'\}$ such that 
$$
(\nexists w_m \in W_U^{\alpha+1}) \,(w_{n_a}R_U^{\alpha+1} w_m)
$$ 
we assign the modal formula $\varphi^i_{n_a} = \varphi^i_a$, where $a \in \mathcal{A}$ is an atomic set, i.e. a set with no internal membership structure being a member of the class $\mathcal A$ of all atoms. In that case any valuation assignment function $v_{n} \in V_U^{\alpha+1}$  with $n = n_a$ returns the value $1$ for any such $\varphi^i_{n_a}$. Then it follows that for all $n \in \{1,\ldots,N'\}$ each valuation assignment function $v_n \in V_U^{\alpha+1}$ gives
\begin{equation}
\label{svamu}
v_n(\varphi_n^i) = 1
\end{equation}
due to the inductive application of the valuation assignment rules for modal formulas, c.f. rules (\ref{nec}) and (\ref{pos}) in section \ref{modal}. Also, for any $m \neq n$ it is $v_m(\varphi_n^i) = 0$, or in other words, the Singleton Valuation Assignment is valid.

\hspace{10 cm}\hfill\boxf\medskip 

The structural unfolding rule, equation (\ref{unfoldrule}), is the first link in a chain of arguments that eventually will result in the description of the aforementioned self-organization of measurements. The rule itself does not have any direct physical meaning; instead this rule and the process that it generates will attain a physical meaning only {\it a posteriori} through what we call a {\it simulation} on a quantum computer representing the physical universe, and in the next paragraphs we will explain this what we mean by a {\it simulation} (we will show that it can naturally be described by the mathematical concept of {\it bisimulation}). Then cosmology, the physics of evolution of the quantum universe, becomes an iterative process on a quantum computer (a certain program, so to say) with one special aim: to perform a simulation of the structural unfolding of $U$. Recalling the principles of Structural Set Theory, the fact is that we cannot have any information about the unfolding's subject prior to its own unfolding. So, given a completely unknown universe $U$, how does the related unfolding process start? First, to put a definite meaning to the unfolding rule, the process' {\it initial} sentence $\varphi^0_U$ has to be specified, because it is from this primitive nucleus where all successive analytical experiments, i.e. the stages of unfolding, follow. Thus before any simulation could ever be executed, the structural unfolding process itself has to be initialized by what we call a {\it semantic realization} of $\varphi^0_U$, which is the point where the initial modal sentence is filled with some meaning.  Considering deeper questions about the nature of the initial sentence as to whether we do have a principle access to its meaning through the scientific method, or even whether we are in the principle position to deduce it from more primitive arguments lies mostly beyond the scope of this work. In the next section at least, we will argue that $\varphi_U^0$ has to be satisfied by a non-wellfounded object.\\
In any case, we outline our original thought again because it will serve us as a guiding principle henceforth:\\[-.2 cm]

{\sf Simulation Principle.} {\it The physical universe is a simulation of the struc\-tural unfolding process of a principally unknown class $U$.}\medskip

With this principle our goal is to find a proper method that allows us to describe the communication between the structural unfolding process and its simulation on the universal quantum computer. As was already motivated in the introduction, the method in question comes with Kripke structures. How can a structural unfolding process of an object $U$ naturally lead to a Kripke structure? In section \ref{baltag} we have seen that at any ordinal stage a Kripke structure can be realized easily. Take $U^{\alpha+1}$ to be a set that satisfies $\varphi^{\alpha+1}_U$ then $U^{\alpha+1}$ can be regarded as the point node of a Kripke structure; and all elements in $U^{\alpha+1}$ become immediate successor nodes of the point if we additionally regard the converse membership $\in^{-1}$ as the directed edge binary relation on all these nodes. But in general the successor nodes themselves are not structureless as they again satisfy some modal sentences generated by our unfolding rule. In this way each successor node contains a secondary Kripke structure in the aforementioned sense. If we iterate this procedure and concurrently identify each node with the corresponding modal sentence that it satisfies, we come up with a number of all modal sentences available at the $\alpha+1$ stage-- which is the number $N'$ in equation (\ref{nst}).\footnote{Here, we have made the assumption that at any ordinal stage the number of modal sentences is always finite, and thus $N'$ remains finite at each stage. This assumption becomes consistent with our basic requirement saying that for each finite ordinal stage we have only a finite quantum universe register.} As a consequence, we obtain a set of all modal sentences at unfolding stage $\alpha+1$ which then becomes the set of possible worlds $W_U^{\alpha+1}$. Finally, equipped with accessibility relation $R_U^{\alpha+1}$ between modal sentences, this set becomes a pointed directed graph. We observe that $R_U^{\alpha+1}$ is not transitive. Within this graph there may be nodes which do not have any internal membership structure anymore. For example, let $U^{2} = \{\emptyset, 1\}$ be an unfolding of the second ordinal stage then the empty set $\emptyset \in U^2$ satisfies some modal sentence $\varphi^1_\emptyset$; and in general many such sentences may occur where each one is satisfied by a member of the class of all atoms $\mathcal{A}$, such as $\emptyset \in \mathcal{A}$. Because of the given satisfaction for atoms, all associated value assignment functions are trivially determined to return a truth value $(T)$; this is simply because satisfaction $a  \models \varphi^i_{n_a},\,\, a \in {\A},$ is equivalent to $v_{n_a}(\varphi^i_{n_a}) = 1$. Having assigned this value to all modal sentences that are satisfied by atoms we may then inductively verify the desired Kripke structure $\kri_U^{\alpha+1}$ through our modal satisfaction rules, c.f. equations (\ref{nec}) and (\ref{pos}).

\subsection{Preferred basis choice \label{preferred choice}} 
We are now ready to introduce the Bisimulation Principle. At any given stage $\alpha+1$ there is a secondary Kripke model  $\kri_\Sigma^{\alpha+1}$ having the following properties
\begin{enumerate}
\item   $\kri^{\alpha+1}_\Sigma = \langle W^{\alpha+1}_\Sigma, P^{\alpha+1}_\Sigma, V_\Sigma^{\alpha+1} \rangle$ is a Kripke structure, with $|W^{\alpha+1}_\Sigma| = |W_U^{\alpha+1}| = N'$, with $P^{\alpha+1}_\Sigma$ being a proximity relation but now acting as an accessibility relation on $W_\Sigma^{\alpha+1}$, and with $V^{\alpha+1}_\Sigma$ being a set of value assignment functions, respectively. 

\item There is a  decomposition $P_\Sigma^{\alpha+1} = \Pp_\Sigma^{\alpha+1} \cup \Pm_\Sigma^{\alpha+1}$ such that for any $\psi \Pp_\Sigma^{\alpha+1} \psi'$ there is exactly one $\psi' \Pm_\Sigma^{\alpha+1}\psi$, i.e. $\Pp_\Sigma^{\alpha+1}$ is the inverse relation to $\Pm_\Sigma^{\alpha+1}$. Then for every $\psi \in W_\Sigma^{\alpha+1}$ there is a unique $\varphi_n^i \in W_U^{\alpha+1}$ such that $\psi$ and $ \varphi_n^i$ are bisimilar, i.e. there exists a {\it bisimulation} $\equiv_B$ with $\psi \equiv_B \varphi_n^i$. All such pairs $\psi\equiv_B \varphi_n^i \in W_\Sigma^{\alpha+1} \times W_U^{\alpha+1}$ are disjoint. By a bismulation we mean binary relation $\equiv_B$ on $W_U^{\alpha+1} \times W_\Sigma^{\alpha+1}$ satisfying
\begin{enumerate}
\item if $\varphi\equiv_B\psi$ then $(\exists j)(v_j(\varphi) = 1) \Leftrightarrow (\exists j')(v_{j'}(\psi) = 1)$,
\item  if $\varphi\equiv_B\psi$ and $\varphi R_U^{\alpha+1}\varphi'$ then $(\exists \psi')(\psi \Pp_\Sigma^{\alpha+1} \psi' \,\,\mbox{and}\,\, \varphi' \equiv_B \psi')$,
\item if $\varphi\equiv_B\psi$ and $\psi \Pp_\Sigma^{\alpha+1} \psi'$ then $(\exists \varphi')(\varphi R_U^{\alpha+1}\varphi' \,\,\mbox{and}\,\,  \varphi' \equiv_B \psi')$,
\end{enumerate}
where $\varphi, \varphi' \in W_U^{\alpha+1}$ and $\psi, \psi' \in W_\Sigma^{\alpha+1}$; and, for convenience, we have used the same symbols for the valuation assignment functions $v_j$, $j \in \{1,\ldots, N'\}$, in both Kripke structures.
\end{enumerate} 

  Let us explain these properties. First, the existence of a Kripke model $\kri_\Sigma^{\alpha+1}$ is postulated in which the proximity relation $P_\Sigma^{\alpha+1}$ plays the role of an accessibility relation. In out understanding. this Kripke model will eventually turn out to be the quantum universe. We know that $\kri_U^{\alpha+1}$ consists of a directed graph, i.e. the relation $R_U^{\alpha+1}$ is not symmetric, while on the other hand $P_\Sigma^{\alpha+1}$ is a symmetric accessibility relation by definition. Thus for the second property to be meaningful, we have to choose only that collection of ordered pairs in $P_\Sigma^{\alpha+1}$ that corresponds to all ordered pairs of possible worlds in $R_U^{\alpha+1}$. Without loss of generality, this choice is made by taking the relation $\Pp_\Sigma^{\alpha+1}$ out of the unique decomposition $P_\Sigma^{\alpha+1} = \Pp_\Sigma^{\alpha+1} \cup \Pm_\Sigma^{\alpha+1}$ which in turn implies a decomposition $\kri_\Sigma^{\alpha+1} = \krip_\Sigma^{\alpha+1} \cup \,^{-}\!\kri_\Sigma^{\alpha+1}$. As a consequence, the intended bisimulation is well posed, and so it shows that two worlds in two different Kripke structures $\kri_U^{\alpha+1}$ and $\krip_\Sigma^{\alpha+1}$ have an equivalent structure in terms of their valuations and in terms of their accessibility relations. Any two bisimilar worlds in Kripke models satisfy the same (infinitary) modal language, i.e. that they are observationally equivalent. And even the converse is true, i.e. two worlds satisfy the same modal language if they are bisimilar (The proof of this equivalence is given in chapter 19 of \cite{mil1990}.). Thus in finite systems observational equivalence and bisimulation are the same although this fact does not need to hold for large systems \citep{bal1999}. Since the second condition further requires that there are exactly $N'$ of disjoint bisimilar pairs between $\kri_U^{\alpha+1}$ and ${^+\!\mathbf M}_\Sigma^{\alpha+1}$, we say that these Kripke structures are bisimilar and write $\kri_U^{\alpha+1} \equiv_B\krip_\Sigma^{\alpha+1}$. In this manner, with bisimulation, we have found a framework to express our preliminary simulation principle with mathematical rigor:\medskip

{\it {\sf Bisimulation Principle.} $\forall \alpha: \,\kri_U^{\alpha} \equiv_B\krip_\Sigma^{\alpha}$}\,.\medskip

Here, we have replaced $\alpha+1$ with $\alpha$ because this principle is regarded from now on as a general postulate for all stages of unfolding-- and thus it stands independent of our ongoing inductive proof.\medskip

Next we quote a result which will be proved at the end of section \ref{metric spaces}. The proof requires a slightly different mathematical context which is the reason why we do not present it right away. Nevertheless,  this result demonstrates already the relevance of the Bisimulation Principle, and so it will help us to derive several conclusions.

\begin{proposition}
\label{bpresults}
The Bisimulation Principle valid at stage $\alpha +1$, {\it viz.}
$$
\kri_U^{\alpha+1} \equiv_B \krip_\Sigma^{\alpha+1}
$$ implies the following statements:
\begin{enumerate}
\renewcommand{\labelenumi}{(\roman{enumi})}
\item There is a Hilbert space $\hil^{\alpha+1} \supseteq \hil^\alpha$ with $\dim \hil^{\alpha +1} = N'$.
\item There is a preferred basis $b^{\alpha+1} \in \basess$, i.e. the basis elements form a family of orthonormal elements in $\hil^{\alpha+1}$ such that each basis element is a non-degenerate eigenvector of a Hermitian operator $\Sigma^{\alpha+1}$ acting on $\hil^{\alpha+1}$.
\item The preferred basis naturally extends to a proximity space  $(b^{\alpha+1}, P_\Sigma^{\alpha+1})$ and, moreover, to the Kripke structure $\kri_\Sigma^{\alpha+1}$.
\end{enumerate}
\end{proposition} 
This proposition completes the inductive proof.\hfill\boxf\medskip

The Bisimulation Principle proposes the existence of a {\it physical space} $\kri_\Sigma^{\alpha+1}$ equipped with a proximity relation $P_\Sigma^{\alpha+1}$, which we identify as the quantum universe. In the previous section we have argued already that the unfolding process itself does not attain a direct physical meaning. Therefore the Kripke structure $\kri_\Sigma^{\alpha+1}$ is the {\it physical place} where the simulation of the structural unfolding happens. In fact, Proposition \ref{bpresults} shows us that this physical place comes with the desired Hilbert space properties. This is one of the cornerstones of the Bisimulation Principle.\medskip

Let us further comment on the properties of the Kripke structure $\kri_\Sigma^{\alpha+1}$.  The main step here has been the identification of a the proximity relation $P_\Sigma^{\alpha+1}$ with the accessibility relation for Kripke structures. In this context the question immediately arises asking for the origin of a proximity relation within a self-experiment of the quantum universe. First of all, as has been argued in section \ref{percep}, all physical experiments--including those experiments which involve the whole universe-- are expected to be imperfect. We think that the origin of the proximity relation really is a basic principle saying that every self-observing system in nature, such as is likely the quantum universe to be, is incapable to perform a self-experiment that would allow to resolve its own underlying physical structure completely. What we mean by {\it underlying physical structure} is that should there be fundamental physical degrees of freedom accessible through physical experiments, then no self-observing system in nature could totally resolve its own state in terms of these degrees of freedom. Also, {\it all} experiments conducted within the universe are self-measurements of the latter, because no experiment conducted in nature can be totally isolated from the surrounding universe. So, at any experiment that the quantum universe organizes onto itself there are at least two (but usually many more) mathematically well distinguished outcomes (or, degrees of freedom) blurred in such a way that they cannot be separated through the outcome of the test. As we know, this fundamental fuzziness is mathematically manifested in the proximity relation. Using our notion of imperfect experiments we make this thought to our second guiding principle.\\[-.2 cm]

{\it {\sf Imperfection Principle.} Every measurement within the universe is imperfect, that is, it gives rise to a non-trivial proximity relation.}\medskip

Evidently, the Imperfection Principle is logically weaker than the Bisimulation Principle in that it only requires the existence of a proximity relation but does not dictate its mapping structure.\medskip

With the Bisimulation Principle we note another insight into the the quantum universe. Recall that for any ordinal $\alpha$ the proximity relation $P_\Sigma^{\alpha+1}$ is {\it reflexive}, i.e. it is $\psi P_\Sigma^{\alpha+1}\psi$ for all $\psi \in W_\Sigma^{\alpha+1}$. This obviously is true because it is $x \in Q_x$ for any quantum $Q_x$, which trivially means that any object is indistinguishable from itself. Therefore, when seen as a node within a labeled graph,  each possible world in the Kripke structure $\kri_\Sigma^{\alpha+1}$ is pictured along with a closed loop representing reflexivity.  Then, since $\kri_U^{\alpha+1} \equiv_B\krip_\Sigma^{\alpha+1}$ holds, the accessibility relation $R_U^{\alpha+1}$ in $\kri_U^{\alpha+1}$ has to be reflexive, too. Since $R_U^{\alpha+1}$ represents the (inverse) membership relation for sets $\in ^{-1}$, this requirement leads us to the proposition that each possible world in $\kri_U^{\alpha+1}$ is satisfied by an  {\it non-wellfounded} set, because we readily have 
$$
\varphi_n^i R_U^{\alpha+1} \varphi_n^i \,\Leftrightarrow \, u_n \in u_n \,,
$$ 
with $n \in \{1,\ldots,N'\}$ and with $u_n \models \varphi_n^i$ for any $\varphi \in W^{\alpha+1}_U$. We cannot further determine whether the objects $u_n$ really are (non-wellfounded) sets in the sense of STS, as they still might be proper classes or large systems; this ignorance, of course, is due to our lack of knowledge about the semantic realization of the initial modal sentence $\varphi_U^0$ standing at the beginning of the structural unfolding process of $U$. Any object satisfying $\varphi_U^0$ must be an atom, i.e. it does not have any successor node in $\kri_U^{\alpha+1}$ except for itself. As a consequence, the non-wellfounded, or, self-referential, character of such an object is inherited into all possible worlds in $\kri_U^{\alpha+1}$ by means of the structural unfolding rule. Thus the Bisimulation principle exposes a feedback from $\krip_\Sigma^{\alpha+1}$ to $\kri_U^{\alpha+1}$ in that it requires that all modal sentences $\varphi_n^i$ in $\kri_U^{\alpha+1}$ must be satisfied by non-wellfounded sets. At this point we quote a result about those sentences \citep{as2004, bal1999}: if a set $a$ is non-wellfounded then it is characterizable only by a sentence of {\it infinitary} modal logic. We immediately conclude that all $\varphi_n^i \in \kri_U^{\alpha+1}$ cannot be sentences of finitary modal logic, hence they involve infinite logical conjunctions. Further, it is worthwhile to remark that in \citep{rkhc1996} the property of reflexivity of an accessibility relation is assigned to self-awareness of an agent represented by the Kripke structure in question. With the Bisimulation Principle we thus are in the position to deduce the self-referential character of the quantum universe on a formal level.

\begin{figure}
\hspace{0.4 cm}\epsfig{file=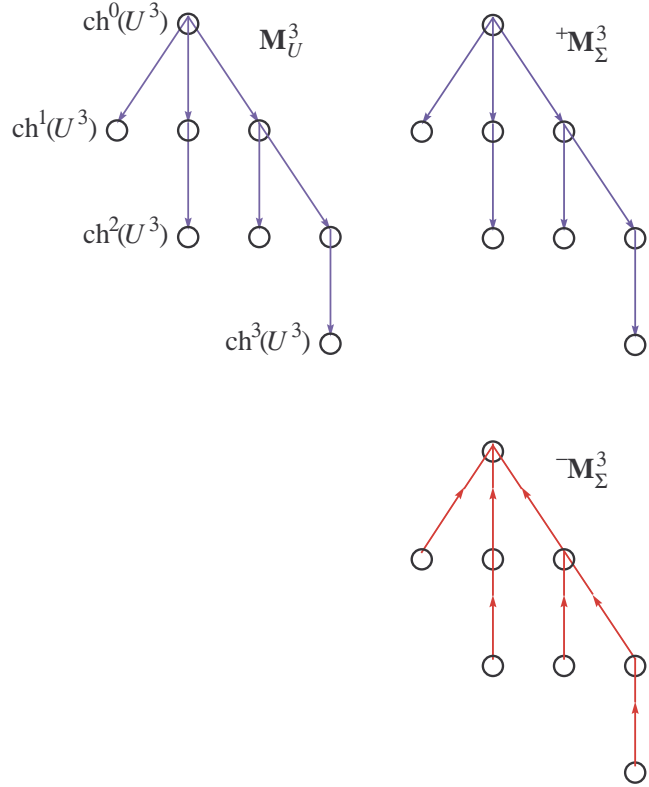, angle=0, width=.78\linewidth}
\caption{Exemplary illustration of the Bisimulation Principle realized at $\alpha = 3$ with $U^3 = \{a_1, 1, 2\} = \{a_1, \,\{a_2\}, \{a_3,\{a_4\}\}\}$, where $a_1, \ldots, a_4 \in {\cal A}$ are atoms. For example, a possible choice would be $a_1 = \ldots = a_4 = \emptyset$ resulting in the von Neumann ordinal $U^3 = 3$. The upper two trees represent the Kripke structures $\kri_U^3$ and $\krip_\Sigma^3$ such that $\kri_U^3 \equiv_B \krip_\Sigma^3$. Arrows in $\kri_U^3$ are the elements in $R_U^3$ representing the inverse membership relation $\in^{-1}$, while the arrows in $\krip_\Sigma^3$ are the ordered pairs of the proximity relation $\Pp_\Sigma^3$. The labels ${\rm ch}^0(U^3), \ldots, {\rm ch}^3(U^3)$ represent sets of children nodes (see,  section \ref{unfoldphys}) such that $Z_U(3) = \sum |{\rm ch}^i(U^3)| = 1 + 3 + 3 + 1 = 8$. The lower right tree is the Kripke model with inverted orientation given through $\Pm_\Sigma^3$. In all trees circles attached to the nodes represent reflexivity of $\Pp_\Sigma^3$ (and of $\Pm_\Sigma^3$) or non-wellfounded objects in $\kri_U^3$, respectively. \label{kripkefig}}
\end{figure}

In figure \ref{kripkefig} we illustrate the Bisimulation Principle employing a simple example.

\subsection{Born's rule \label{borns rule}}
We let the Kripke structure $\kri_U^{\alpha+1}$ be equipped with a weighting function on all possible worlds, $\omega^{\alpha+1}_\cdot: \{1,\ldots,N'\} \rightarrow [0,1]$. Thus given a preferred basis $b^{\alpha+1}$, we may evaluate the basic probability assignment $m(\varphi^i_n)$ for an arbitrary chosen modal sentence $\varphi^i_n \in W_U^{\alpha+1}$.  Due to the fact that the Singleton Valuation Assignment (SVA) is valid in our construction, i.e. exactly one modal sentence is true for each one of the possible worlds in $W_U^{\alpha+1}$ (see equation (\ref{svamu})), we are allowed to evaluate $m$ according to equation (\ref{bpa}),
\begin{eqnarray}
\nonumber
m(\varphi_n^i) &=& \sum_{j=1}^{N'} \omega_j^{\alpha+1} \, v_j\!\!\left(\Box \!\!\left[ \bigvee_{\varphi_n^i R_U^{\alpha+1}\varphi_m^{i-1}} \!\!\varphi_m^{i-1}\right] \wedge \left[ \bigwedge_{\varphi_n^i R_U^{\alpha+1}\varphi_m^{i-1}} \!\!\Diamond \varphi_m^{i-1}\right ]\right )\\
&=&  \sum_{j=1}^{N'} \omega_j^{\alpha+1} \, v_j(\bigtriangleup \, \{\varphi_m^{i-1} : \varphi_n^i R_U^{\alpha+1}\varphi_m^{i-1}\})\\
\nonumber
&=& \omega_n^{\alpha+1}\,,
\end{eqnarray}
which means that only one possible world labeled with $j=n$ yields a truth value of 1 for $\varphi_n^i$ and is therefore weighted with the factor $\omega_n^{\alpha+1}$; in this way the latter turns out to be the basic probability assignment of an arbitrary  modal sentence $\varphi_n^i$. Let $\psi_n \in b^{\alpha+1}$ be the unique basis element that is bisimilar to $\varphi^i_n$, i.e. $\psi_n \equiv_B \varphi^i_n$. We then identify Born's rule as 
\begin{equation}
\label{born}
\omega^{\alpha+1}_n = |(\psi_n \in b^{\alpha+1}, \Psi_\alpha)|^2  = P(\psi_n \in b^{\alpha+1}| \Psi_\alpha)\,, 
\end{equation}
where $\Psi_\alpha$ is the state of the universe at the present stage $\alpha$. \boxf\medskip

\subsection{State selection, prediction, and explanation \label{state selection}}
Given a preferred basis $b^{\alpha+1}$, let $ \psi^* \in b^{\alpha+1}$ be the unique state that is bisimilar to the point $\varphi^{\alpha+1}_{N'} \in \kri_U^{\alpha+1}$,   
then $\Psi_{\alpha+1} := \psi^*$ becomes the state of the universe in the $\alpha+1$ stage.\boxf\medskip

State selection is accomplished here through the rule saying that at every ordinal stage the one preferred basis element becomes the state of the universe that is bisimilar to point, i.e. to the distinguished world, in $ \kri_U^{\alpha+1}$. Obviously, no wave collapse or state reduction takes place; only a given preferred basis element in $\hil^\alpha$ is replaced by another preferred basis element in $\hil^{\alpha+1}$. But why do we select $\psi^*$ element as the new state of the universe? We think that this kind of selection is natural, because $\psi^*$ is bisimilar to the distinguished modal sentence $\varphi_{N'}^{\alpha+1} \in {\mathbf M}_U^{\alpha+1}$  that has been recursively created out of all preceding stages of unfolding. This modal sentence is logically consistent with all $\varphi_U^\beta$ with $\beta < \alpha +1$, and is therefore the one carrying maximum information about the membership structure of $U$ available at stage $\alpha +1$.\smallskip

Even though the above rule describes {\it how} the next state of universe is to be selected, it is inaccessible for any self-referential observer, including the whole quantum universe itself, to determine {\it which} element of the forthcoming preferred basis $b^{\alpha+1}$  will be the next state. This ignorance is a direct consequence of the Imperfection Principle, which says that the structure of the proximity relation $P_\Sigma^{\alpha+1}$ is principally unknown to any self-referential observer, and there is no way for the latter to translate the state selection rule into a deterministic prediction for the next state of the universe $\Psi_{\alpha+1}$. Therefore, we do not have a conflict with the indeterministic nature of quantum physics.\bigskip

We see that what at most can be done in predicting the successive state of the universe is to apply Born's rule-- resulting in a purely probabilistic forecast. This kind of prediction, however, prerequisites a knowledge of the preferred basis $b^{\alpha+1}$ beforehand, and first of all it is questionable whether the preferred basis is at the observer's disposal before the quantum universe completed its state selection step $\Psi_\alpha \mapsto \Psi_{\alpha+1}$. But even if it would be accessible {\it a priori}, then no observer could ever acquire complete information about $b^{\alpha+1}$ because the proximity relation $P^{\alpha+1}_\Sigma$ would cause many elements in $b^{\alpha+1}$ to be perceptually indistinguishable (recall that the existence of preferred basis already implies that there is a proximity relation on it-- as stated in Proposition \ref{bpresults}). We therefore conclude that Born's rule in its usual form, equation (\ref{born}), simply does not represent correctly the situation which an observer within the quantum universe is confronted with. Instead, a multi-valued version of Born's rule-- taking into account the proximity relation-- has to be introduced such that point-like indiscernible elements of $b^{\alpha+1}$ in equation (\ref{born}) are replaced with quanta. Thus ``Born's rule'' in this case reads
\begin{eqnarray}
\label{bornbel}
 P^*(\psi \in b^{\alpha+1} | \Psi_\alpha) &:=& \sum_{\varphi^i_n \in Q'} m(\varphi^i_n) =  \sum_{\psi' \in Q}|(\psi' \in b^{\alpha+1}, \Psi_{\alpha})|^2 \,, 
\end{eqnarray}
where the quantum set $Q \in {\mathcal Q}_P(b^{\alpha+1})$ is defined as the union of all quanta that contain $\psi$ as an element, and where $Q'$ is the collection of modal sentences, i.e. possible worlds in $\kri_U^{\alpha+1}$,  such that $Q' \equiv_B Q$. Equation (\ref{bornbel}) may directly be rewritten in terms of a belief measure Bel on ${\mathcal P}(b^{\alpha+1})$. For this purpose define the basic probability assignment $m:{\mathcal P}(b^{\alpha+1}) \rightarrow [0,1]$ as
$$
m(A) := \left\{ \begin{array}{lll}
m(\varphi_n^i) & \mbox{if} \, A = \{\psi\} \,\, \mbox{with} \,\, \varphi_n^i \equiv_B \psi\,,&\\
0 &  \mbox{otherwise.} &
\end{array}\right.
$$
Then equation (\ref{bornbel}) reads as
\begin{equation}
\label{bornbel2}
 P^*(\psi \in b^{\alpha+1} | \Psi_\alpha) = \sum_{A \subseteq Q} m(A) = \bel(Q)\,.
\end{equation}
It is worthwhile to emphasize that our construction of this belief measure effectively restricts the domain of Bel to the complete ortholattice $\orthop$ of quantum sets in ${\mathcal P}(b^{\alpha+1})$. Therefore a value $\bel(Q)$,  with $Q \in {\mathcal P}(b^{\alpha+1})$, is defined only if $Q$ is a quantum set. This restriction stems directly from the fact that we always look for all quanta that have our element of concern as a member; only on this union of relevant quanta the belief measure Bel may be evaluated.\footnote{We  may easily enhance to the domain of Bel to the whole power set ${\mathcal P}(b^{\alpha+1})$, if for {\it any} $A \in {\mathcal P}(b^{\alpha+1})$ we take $\bel(A) := \bel(Q_A)$, where $Q_A$ is the smallest quantum set that contains $A$ as a subset, but this enhancement does not provide any new insights.} In other words, traditional Born's rule, equation (\ref{born}), induces a probability measure on ${\mathcal P}(b^{\alpha+1})$ interpreted as a $\sigma$-algebra of events ${\mathcal P}(b^{\alpha+1})$. On the contrary, due to equation (\ref{bornbel2}),  a given proximity relation $P_\Sigma^{\alpha+1}$ induces a belief measure Bel on the set of all quantum sets ${\mathcal Q}_P(b^{\alpha+1}) \subseteq {\mathcal P}(b^{\alpha+1})$ interpreted as a complete ortholattice of events $\orthop = ({\mathcal Q}_P(b^{\alpha+1}), \,\cap_P, \,\cup_P, \,^\perp)$.\bigskip
 
After state selection, the universe is in the state $\Psi_{\alpha+1}$. At this moment of exotime we may ask for a {\it probabilistic explanation} of the present state (In the same way as Born's rule gives a {\it probabilistic prediction} for possible next states). We ask: given $\Psi_{\alpha+1}$, what is the probability assignment for all preferred basis elements in $b^{\alpha}$? But we may not apply Born's rule again in order to explain the present state $\Psi_{\alpha+1}$, because this choice will not lead to a probability measure on ${\mathcal P}(b^\alpha)$ in general, {\it viz.}
$$
\sum_{i=1}^{N} |(\Psi_{\alpha+1}, \psi_i \in b^\alpha)|^2 < 1 \,.
$$
Probabilistic explanation (often referred to as {\it conditioning}) usually  is calculated with the Bayes rule. According to it the {\it posterior probability} $P(\psi \in b^\alpha| \Psi_{\alpha+1})$  at stage $\alpha +1$ reads as
\begin{equation}
P(\psi \in b^\alpha| \Psi_{\alpha+1}) = \frac{P( \Psi_{\alpha+1} | \psi \in b^{\alpha+1}) \, P(\psi \in b^{\alpha})}{P(\Psi_{\alpha+1})} \,,
\end{equation}
where $P(\psi \in b^{\alpha})$ is the {\it prior probability} of a state in $\hil^{\alpha}$ realized as $\psi$, and where the normalizing constant is (we write simply ``$\psi$'' instead of ``$ \psi \in b^\alpha$'')
\begin{eqnarray}
\nonumber
P(\Psi_{\alpha+1}) &=& P( \Psi_{\alpha+1} | \psi) \, P(\psi) + P( \Psi_{\alpha+1} | \neg \psi) \, P(\neg \psi)\\
\nonumber
&=& |(\Psi_{\alpha+1}, \psi)|^2 \, P(\psi) + (1 - P(\psi))\sum_{\psi' \perp \psi} |(\Psi_{\alpha+1}, \psi')|^2 \,.
\end{eqnarray}
A calculation of the posterior prerequisites a knowledge about the value of the prior. But, like in the case of prediction, no observer present in stage $\alpha+1$ is able to tell what basis element in $b^\alpha$ used to be the exact state of the universe, because likewise there are at least as many alternatives possible as there are elements in the associated quanta. We are again confronted with multivalued mappings so that a treatment of this problem in the sense of Dempster-Shafer theory is indicated. Thus we may replace equation (\ref{bornbel}) with its associated value of the belief measure Bel on ${\mathcal Q}_P(b^\alpha) \subseteq {\mathcal P}(b^{\alpha})$:
\begin{equation}
P^*(\psi \in b^\alpha| \Psi_{\alpha+1}) := \sum_{A \subseteq Q} m(A) = \bel(Q)\,,
\end{equation}  
where $Q$ is the quantum set being the union of of all quanta (generated through the proximity relation $P_\Sigma^\alpha$) that hold $\psi \in b^{\alpha}$ as a member, and where $m: {\mathcal P}(b^{\alpha}) \rightarrow [0,1]$ is defined as
$$
m(A) := \left\{ \begin{array}{lll}
P(\psi \in b^\alpha | \Psi_{\alpha+1}) & \mbox{if} \, A = \{\psi\} \,\, \mbox{with} \,\, \varphi_n^i \equiv_B \psi\,,&\\
0 &  \mbox{otherwise,} &
\end{array}\right.
$$
with $\varphi^i_n \in {\mathbf M}_U^{\alpha}$.\medskip

In summary, we have shown that any observer-- including the quantum universe itself-- can make quantitative statements about the future or about the past in exotime only in terms of modalities representing the degree of belief, according to Dempster-Shafer theory of evidence. This is the main difference to the ``traditional'' approach via Born's rule.

\section{Metrics, tree metrics, and embeddings\label{metric spaces}}

A basic test for any quantum mechanical description of the universe is the necessity for an explanation of an apparently smooth three-dimensional manifold structure that on many length scales does not exhibit any quantum character whatsoever. A smooth three-dimensional space is one of the basic pillars of our external experience. Surely, there are further levels of difficulties related to this issue; for instance, the problem of how a quantum treatment of the universe may plausibly emerge into a unified description of space {\it and} time resulting in a four-dimensional manifold structure being locally isomorphic to Minkowski space. And finally, there still remains the open question of how a general representation of space, time {\it and} matter could ever be accomplished in such an approach to incorporate full General Relativity. With regard to a solution of these problems we have just gained first insights; so, for example, Eakins and Jaroszkiewicz \citep{ej2003, ej2004a} propose that the factor structure of the selected state $\Psi_\alpha$ of the universe may ultimately be responsible for a classical Einstein universe. In their interpretation, the interplay of factorized and entangled states may give rise to causal sets, i.e. to the basic building blocks of Einstein locality.\medskip

The present work permits for a slightly different point of view on the problem of the basic building blocks of the universe, i.e. the fundamental degrees of freedom. Our understanding is that at present there are at least two different  categories of approaches to this problem.  The first category contains all those attempts which recognize one paradigm, namely, that the fundamental degrees of freedom of the universe must be closely related to the set of elementary degrees of freedom in General Relativity, that is, to geometrical points on a Lorentzian four-manifold. All attempts that try to construct a quantization of General Relativity certainly fall into this category.  But there is a second category in which it is not presumed {\it a priori} that such a relation to General Relativity exists. Physical approaches of the second type look for other--but not less significant-- aspects of nature that are not directly associated with relativistic space-time structure. Surely, in a certain approximation or limit these approaches have to prove consistency with the principles of General Relativity but there is no necessity to explain the intended consistency in the first place. We think that the description of the self-referential quantum universe presented in this work could be a candidate for the second category. We will show that this description will provide us with a collection of degrees of freedom being of quite a different nature than a collection of geometrical points constituting a smooth manifold. Thus although a consistency proof with General Relativity remains to be done, we put forward the hypothesis that there are other aspects of nature that any theory of the elementary degrees of freedom in physics has to meet. The goal of this section is to explore these aspects and to identify their degrees of freedom.\medskip

Following Bell \citep{bel2000, bel1986}, we introduce the notion of continuity for proximity spaces. Given a proximity space $(X, P)$ we say $(X, P)$ is {\it $P$-continuous} if for any $x,y \in X$ there is a set $\{x,z_1, \ldots, z_n,y\}$  such that the set $\{xPz_1, \ldots, z_nPy\}$ exists.  We conclude that even if $X$ is discrete, a proper notion of perceptual continuity can be defined because within each sequential pair of points in  $\{x,z_1, \ldots, z_n,y\}$ one point is indiscernible from the other. We call the set  $\{x,z_1, \ldots, z_n,y\}$ {\it open path} from $x$ to $y$ and concurrently assume that an open path does not contain closed paths, i.e. each element in $\{x,z_1, \ldots, z_n,y\}$ appears exactly once within the open path. We define the length of an open path as $l(x,y) := |\{x,z_1, \ldots, z_n,y\}| -1$, and set the trivial case $l(x,x) = 0$ for all $x,y \in X$. For any ordinal stage of the quantum universe consider now the Kripke structure $\kri_\Sigma^\alpha$. The set $\kri_\Sigma^\alpha$ is a proximity space (as it is equipped with the associated proximity relation $P_\Sigma^\alpha$), and it also is $P$-continuous. Even more, $\kri_\Sigma^\alpha$ is a {\it tree}, because for any $\psi, \psi'$, with $\psi \neq \psi'$, there is exactly one path leading $\psi$ to $\psi'$. The tree property of $\kri_\Sigma^\alpha$ directly follows from the fact that $\kri_U^\alpha$ is constructed as a tree, and from the validity of the bisimulation principle $\kri_U^\alpha \equiv_B \krip_\Sigma^\alpha$. We employ the uniqueness of open paths in $\kri_\Sigma^\alpha$ to define the discrete {\it tree metric} $d_T^\alpha$ on $\kri_\Sigma^\alpha$ as 
$$
d_T^\alpha: W_\Sigma^\alpha \times W_\Sigma^\alpha \rightarrow {\mathbb N}_0\quad \mbox{with}\quad d_T(\psi,\psi') := l(\psi,\psi').$$ 

We find that it is again the Bisimulation Principle which invokes this discrete metric structure on $\kri_\Sigma^\alpha = b^\alpha \subset \hil^\alpha$. But what does this metric mean physically? If we consider two separated sets $\{\psi\}, \{\psi'\} \subset b^\alpha$ in the sense of separation given in section \ref{percep}, then the two corresponding elements in $\hil^\alpha$ are perceptually (or by means of any self-experiment in the universe) distinguishable, because there is no $\psi P_\Sigma^\alpha \psi'$. From a physical point of view these elements represent two objects of the universe which should exhibit a quantitative similarity relation mathematically equivalent to the tree metric $d_T^\alpha$. Before we explore further the mathematical and physical implications of the tree metric, let us briefly reconcile the general character of metrics in physics, and here especially the role of distance in space.\medskip

In physics, the elementary similarity relation between two objects normally is their distance in three-dimensional space. It is given by a value of a function $d$ conventionally understood as a metric on a three-dimensional Riemannian manifold $M$.  Distance in space has always been seen as the most fundamental mathematical relation in physics, because space itself has been understood as the stage where all physical action happens. Before the advent of General Relativity space had the role of a completely rigid and passive structure unable to expose any interaction or feedback with physical objects. Space (and time) served solely as a mathematical configuration space (sometimes also called a {\it block universe})-- not more than a convenient labeling method for physical objects in coordinates of three-space and in time. General Relativity gave space and time a dynamical role and therefore a true physical meaning. However, General Relativity still shares the point of view that (local) three-dimensional space and time ought to be fundamental elements of physical experience. This heritage is a remainder from times when the universe was regarded as a rigid bock and it finds its expression in the fact that Einstein's field equations determine a metric tensor of a four-manifold as a solution.  But what quantum theory taught us among many other things is the important lesson that physical objects often have degrees of freedom that in general do not admit a proper description in a configuration space being a smooth three-dimensional Riemannian manifold (We do not consider time as true physical degree of freedom henceforth.). The quest for a theory of quantum gravity is the search for a theory of the fundamental degrees of freedom in physics. We therefore believe that difficulties must arise in any attempt to construct a quantum version of General Relativity, simply because the latter initially narrows the view to three-space as a candidate for a fundamental configuration space in physics while the former allows for a broader view where the elementary degrees of freedom might well belong to a completely different configuration space. Therefore, it is at least questionable why the elementary degrees of freedom in physics should  form the domain of a metric in three-space. Surely, there should be a proper limit in which a metric in three-space could be recovered, but this requirement does not invalidate the previous argument. Still, fundamental degrees of freedom in physics must nevertheless exhibit the possibility of pairwise comparison by means of a mathematical relation that is physically plausible at the same time. This assumption is reasonable because in any physical theory there must be an option to decide whether two accessible degrees of freedom can be distinguished or not, and further, there should also be a plausible degree of similarity for already distinguished degrees of freedom.\medskip  

In our approach such a similarity relation naturally comes with the tree metric $d_T^\alpha$ operating on the Kripke structure $\kri_\Sigma^\alpha$. So, with the previous arguments in mind, we want to extend our view on similarity relations in physics and ask: are there physical objects that are comparable by means of a tree metric? We think that there are such objects and in order to find them we have to recall a fact, namely, that physical objects are carriers of information. This means that in general a physical object's identity not only consists of the assembly of its physical constituents,  but that an inherent part of the object's identity can be found only within the amount of information that it concurrently carries. Therefore, a reductionist representation of an object in question by means of its conventional physical building blocks would often not reveal the object's true identity in nature. To illustrate this issue, let us give three examples. A printed book, for instance, could be correctly described by means of a vast amount of individual physical particles altogether forming a certain solid state. Such a description would involve a gigantic collection of equations representing the fabric of the paper, while other equations would describe the behavior of ink particles, and so on. However, such a representation would make it practically impossible to decipher the content of the book's story and therefore an immanent part of the book's identity would be lost. Another example are black holes. Black holes {\it can} be interpreted as classical solutions of Einstein's equations but we expect that this is probably only a small part of the whole story. In the past three decades there has been a growing evidence that black holes are carriers of information placed on their surface, i.e. on their event horizon. This information is  likely to be accessible only through a correct quantum description of black holes. A classical black hole solution in terms of General Relativity thus is at most a simple description of the physical carrier but it is certainly not a suitable representation of the information it--the black hole-- actually holds. Our third example is the biological macromolecule DNA. Here too we may give a reductionist description giving rise to a vast collection of interacting atoms in three spatial dimensions but such a representation would again hardly reveal something about the biological meaning of the genetic code. Hence, it is not hard to find physical objects which are {\it primely} carriers of information and which only play a secondary role as extended objects in space. Thus, any theory which aims to represent the true physical degrees of freedom in our universe has to address this issue. For instance, it would not make much sense to compare two books, say, by their spatial distance  in order to describe the information similarity of their contents-- so, locations in space simple cannot be the only fundamental degrees of freedom in nature. Other similarity relations with different domains are therefore necessary between physical carriers of information. And indeed, as we are now ready to show, tree metrics and related structures turn out to be very useful when it comes to compare the {\it information} encoded in physical objects.\medskip

Since Shannon's theory of communication \citep{sha1948} it is known that any kind of information can be stored in symbolic sequences. Symbolic sequences consist of symbols which are taken from  a (finite) alphabet. Any (english) book consists of a sequence of letters from the (english) alphabet. In a DNA we have sequences of chemical symbols from the alphabet $\{A, C, G, T \}$. And even if we still do not know the information code of black holes, an opinion assuming that the event horizon is imprinted with letters from an hypothetical alphabet does at least not contradict our present understanding of these mysterious objects. Further, Shannon has shown that any symbolic sequence can be translated into a mathematically equivalent sequence built from zeros and ones only, i.e. into a sequence from the binary alphabet $\{0,1\}$. Could it therefore be-- and this is the main motivation for the remainder of this section-- that even the fundamental degrees of freedom in the quantum universe are symbolic sequences? 

We want to discuss this problem at a basic level using results from the mathematical theory of {\it cut polyhedra} (For this purpose we follow chapter 3 from the review article of Deza and Laurent \cite{dl1994}). One of the immediate applications of this theory is the problem of embedding discrete metrics into normed spaces. Our aim is to look for a canonical {\it isometric} embedding of the tree metric $d_T^\alpha$ (that is a distance preserving embedding) into a normed space. We begin with a few preliminaries. Let $l_p$ be the space of sequences of real numbers with norm
$$
\|x\|_p : = \left( \sum_{i=0}^\infty | x_i |^p \right)^{1/p} \,.
$$    
A measure space $(\Omega, {\A}, \mu)$ consists of a set $\Omega$, a $\sigma$-algebra $\A$ of subsets of $\Omega$, and a measure $\mu$ defined on $\A$ which is additive, i.e. 
$$
\mu(\bigcup_{n\geq1} A_n) = \sum_{n\geq1} \mu(A_n)
$$
for all pairwise disjoint sets $A_n \in {\A}$, and satisfies $\mu(\emptyset) = 0$. The measure space is non negative if $\mu(A) \geq 0$ for all $A \in {\A}$. A probability space is a non negative measure space with total measure $\mu(\Omega) = 1$. Given a measure space $(\Omega, \A, \mu)$ and a given function $f: \Omega \rightarrow {\mathbb R}$ , its $L_p$ norm is defined by:
$$
\| f \|_p : = \left(\int_\Omega |f(\omega)|^p \,\mu(d\omega) \right)^{1/p}\,.
$$
Then $L_p(\Omega, \A, \,u)$ defines the set of measurable functions, and the $L_p$-norm defines a metric structure on on $L_p(\Omega, \A, \mu)$.

Given any non negative measure space $(\Omega, \A, \mu)$, another metric space $(\A_\mu, d_\mu)$ can be defined, where $\A_\mu = \{A \in \A: \mu(A) < \infty \}$ and $d_\mu(A, B) = \mu(A \Delta B)$ for $A, B \in \A_\mu$ (here, $A\Delta B = \{a \in A: a \notin B\} \cup \{b \in B: b \notin A\}$ for $A, B \subseteq \Omega$). In fact, $(\A_\mu, d_\mu)$ is the subspace of $L_1(\Omega, \A, \mu)$ consisting of its $\{0,1\}$-valued, i.e. binary, functions. 
 A metric space $(X, d)$ is $L_1${\it-embeddable} if it is a subspace of some $L_1(\Omega, \A, \mu)$ for some non negative measure space, i.e. there is a map $\phi: X \rightarrow L_1(\Omega, \A, \mu)$ such that 
$$
d(x,y) = \| \phi(x) - \phi(y) \|_1 
$$
for all $x,y \in X$. A useful result is the following Lemma (c.f., Lemma 3.5 in \cite{dl1994}).
\begin{lemma}
For a metric space $(X, d)$ the following assertions are equivalent
\begin{enumerate}
\renewcommand{\labelenumi}{(\roman{enumi})}
\item $(X,d)$ is $L_1$-embeddable.
\item $(X,d)$ is a subspace of $(\A_\mu, d_\mu)$ for some non negative measure space $L_1(\Omega, \A, \mu)$.
\end{enumerate}
\end{lemma}
If we take $(X,d) = (M_\Sigma^\alpha, d_T^\alpha)$ then this Lemma guarantees that there is a subspace of $\{0,1\}$-valued and $\mu$-measurable functions containing $(M_\Sigma^\alpha, d_T^\alpha)$ as an isometric embedding. Thus we have established already a connection between binary sequences and $L_1$-embeddable metrics. A standard result is that any finite tree metric $(X,d)$ is isometrically embeddable into a finite subspace of $l_1$, i.e. there exist $n = |X|$ vectors $x_1, \ldots , x_n \in {\mathbb R}^m$ for some $m \in {\mathbb N}$ such that $d_{ij} = \|x_i - x_j \|_1$ for $1 \leq i < j \leq n$. To this result the following theorem immediately applies (c.f., Theorem 3.8 in \citep{dl1994}).
\begin{theorem}
\label{equivl1}
Let $(X,d)$ be a metric space. The following assertions are equivalent
\begin{enumerate}
\renewcommand{\labelenumi}{(\roman{enumi})}
\item $(X,d)$ is $L_1$-embeddable.
\item $(X,d)$ is $l_1$-embeddable. 
\end{enumerate} 
\end{theorem}
We therefore obtain the following Corollary readily.
\begin{corollary}
\label{coll1}
Let $ (W_\Sigma^\alpha, d_T^\alpha)$ be the finite tree metric as defined above with $| W_\Sigma^\alpha| = N$. Then there is a non negative space $L_1(\Omega, \A, \mu)$ of measurable functions and a subspace $(\A_\mu, d_\mu) \subset L_1(\Omega, \A, \mu)$ of binary valued mappings such that $(W_\Sigma^\alpha, d_T^\alpha) \subseteq (\A_\mu, d_\mu)$. 
\end{corollary}
Thus indeed the possible worlds in $W_\Sigma^\alpha = b^\alpha$ are the characteristic functions on $\Omega$. So our original idea-- that symbolic sequences of encoded information may be the building blocks of the quantum universe--  turned out to be correct in some sense. And in order to stronger emphasize the relation to Shannon's theory of communication we give a few additionally remarks. The $d_\mu$-metric is the Hamming metric in $(\A_\mu, d_\mu)$ which counts the number of not equally valued positions between two arbitrary characteristic functions in $(\A_\mu, d_\mu)$. The values correspond to all individual sets $\{q\}$ with $q \in \Omega$. For now we presume that $\Omega$  has finite cardinality $n$, thus any characteristic function in  $(\A_\mu, d_\mu)$ has  finite domain. Let further $C_i \in (\A_\mu, d_\mu)$ denote the elements of the isometric embedding of $(W_\Sigma^\alpha, d^\alpha_T)$, that is, for every $\psi, \psi' \in W_\Sigma^\alpha$ there are indexes $1 \leq i,j \leq N$ such that $\|C_i - C_j\|_1 = d^\alpha_T(\psi, \psi')$.  Can we reach a further characterization of these sequences? One possibility is to regard each $C_i$ as a {\it codeword}. In this manner a code sequence $C_\mu(n,k)$ of $k \in {\mathbb N}$ codewords each of length $n \in {\mathbb N}$ can be encoded as a $k$-tuple
$$
C_\mu(n,k) = (C_{i_1}, \ldots, C_{i_k}) \,
$$ 
for all $i_j \in \{1,\ldots, N\}$ with $1 \leq j \leq k$. We stress that while in Shannon's theory of communication information is encoded solely by means of a binary sequence of values from ${\mathbb Z}_2$ (or from any other finite field), our encoding additionally refers to the associated $\mu$-measurable sets $\supp(C_{i_j}) \subseteq \Omega$ in order to define a binary code properly (Here, $\supp(C_{i_j})$ is the support of $C_{i_j}$, i.e. $\supp(C_{i_j}) = \{e \in \Omega: C_{i_j}(e) = 1 \}$.). Here, the set $\Omega$ can regarded as the carrier of encoded information.

A distinguished class of codes depicts those which exhibit error correction. Suppose a code sequence $C_\mu(n,k)$ has been encoded and transmitted through a noisy channel. After transmission of the first $n$ bits the receiver holds a vector $V_1 \in {\mathbb Z}_2^n$; the receiver also knows all codewords $C_i \in (\A_\mu, d_\mu)$  with $1 \leq i \leq N$. Additionally, we presume that the receiver has the measure $\mu$ on $\Omega$ at hand. The problem of error correction is this: when $V_1 \neq C_{i_1}$, can the receiver recover the correct codeword $C_{i_1}$ from $V_1$? He can, if the sender knows beforehand that during transmission there will be at most $e \in \mathbb{N}$ errors, i.e. $d_\mu(C_{i_1}, V_1) \leq e$ holds,  and if any ball in $L_1$ of radius $e$ contains at most one codeword. In this situation the receiver just has to look for the nearest neighbor of $V_1$ in $L_1$ to get the correct codeword $C_{i_1}$ (this strategy is called the nearest neighbor algorithm.). If the mentioned prerequisites are met then $C_\mu(n,k)$  is an instance of a so-called  {\it error correcting} $[n', N, e]$-code, where $n' = nk$ is the length of the code. There is a condition on the metric structure controlling the ability of error correction. Let $d_m$ be the minimum $d_\mu$-distance of all disjoint codeword pairs then it is straightforward to show that $e \leq (d_m-1)/2$. Since the tree metric $d_T^\alpha$ embeds in $L_1$ isometrically, we have $d_m = 1$, and so it  follows $e = 0$; therefore, $C_\mu(n,k)$ is characterized as a representative of a non error correcting $[n', N, 0]$-code.

\medskip

We have found an appropriate $L_1$-embedding and a $l_1$-embedding of the tree metric $(W_\Sigma^\alpha, d^\alpha_T)$, and we have recognized the elements of this embedding as codewords of a binary code we have not yet identified the elements in $W_\Sigma^\alpha$ as vectors in Hilbert space. But this step can be done readily because the embedding is non negative, i.e. the embedded functions are non negative elements in a non negative measure space $L_1(\Omega, \A, \mu)$. By Theorem \ref{equivl1}, these functions equivalently determine non negative sequences in $l_1$, and every vector space element in $l_1$ of the embedding must be an element of $l_2$, the Hilbert space. This follows because for any non negative sequence in $x \in l_1$ it is $\|x\|_2 \leq \|x\|_1$.\medskip

Let  $V_2 = \{x_1, \ldots, x_N \}$ be the family of vectors in $l_2$ which constitute an isometric embedding of $(W_\Sigma^\alpha, d^\alpha_T)$ in $l_1$. With $V_2$ we define the {\it Euclidean distance matrix} $D^\alpha_2 \in {\rm Mat}({\mathbb C}, N)$ with symmetric matrix entries
$$
(D_2^\alpha)_{ij} = \|x_i - x_j\|_2\,,
$$
for all $v_i, v_j \in V_2$ and $1 \leq i,j \leq N$. Regarding $D^\alpha_2$ as an $N$-dimensional vector space homeomorphism $D_2^\alpha:{\mathbb C}^N \rightarrow {\mathbb C}^N$, we pose the eigenvalue problem
\begin{equation}
(D^\alpha_2  - \lambda )u = 0 \,,
\end{equation}
with $\lambda \in {\mathbb C}$ and $u \in {\mathbb C}^N$. Since $D^\alpha_2$ is symmetric there is a unique orthonormal basis $\{\psi_1, \ldots, \psi_N\}$ of eigenvectors (after a possible normalization) corresponding to an ordered sequence of real eigenvalues $\lambda_1 > \lambda_2 > \ldots > \lambda_N$. This last property together with the fact that $\lambda_1$ must be the only positive eigenvalue of $D^\alpha_2$ are well known results about Euclidean distance matrices due to Schoenberg \citep{sch1937}. Euclidean distance matrices for large $N$ have interesting properties; for example, in \citep{bbc2003} it is shown that when going from small negative to large negative eigenvalues the corresponding eigenvectors experience a localization-delocalization transition. Also, large negative eigenvalues permit a continuous treatment of the Euclidean distance matrix leading to an integral equation being very similar to a Laplace equation.\medskip

In basis of the eigenvectors $\{\psi_1, \ldots, \psi_N\}$ we recover the Hermitian operator
$$
\Sigma^\alpha = \sum_{i=1}^N \lambda_i {\mathsf P}_i
$$
where ${\mathsf P}_i = {\mathsf P}_i^2$ is the projection onto the one-dimensional subspace spanned by the eigenvector $\psi_i$. Then the preferred basis at stage $\alpha$ is
$$
b^\alpha = \{\psi_1, \ldots, \psi_N\} = W_\Sigma^\alpha
$$
spanning the complex Hilbert space
$$
\hil^\alpha = {\rm span}_{\mathbb C} \{\psi_1, \ldots, \psi_N \} = ({\mathbb C}^N, \| . \|_2) \,,
$$
where $\|.\|_2$ denotes the $l_2$-norm for complex vectors which, as usual, is given by the square root of the inner product $(.,.)$ of the Hilbert space $l_2$.\medskip

We are now in the position to prove Proposition \ref{bpresults}. Since the Bisimulation Principle holds at all ordinal stages $\alpha$ it is valid at $\alpha+1$. In this manner we repeat all steps presented in this section with $\alpha+1$ instead of $\alpha$; thus we obtain the Hilbert space 
$$
\hil^{\alpha+1} = {\rm span}_{\mathbb C} \{\psi_1, \ldots, \psi_{N'} \} = ({\mathbb C}^{N'}, \| . \|_2) \,,
$$
which has $\dim \hil^{\alpha+1} = N'$ and which has $\hil^\alpha$ as a subspace. The family of eigenvectors $\{\psi_1, \ldots, \psi_{N'}\}$ is the preferred basis in $\hil^{\alpha+1}$, i.e.
$$
b^{\alpha+1} := \{\psi_1, \ldots, \psi_{N'}\}\,,
$$
and the Hermitian operator $\Sigma^{\alpha+1} = \sum_i \lambda_i {\mathsf P}_i$ is non-degenerate. This proves the points {\it (i)} and {\it (ii)} in \ref{bpresults}. To show point {\it (iii)} it suffices to consider $b^{\alpha+1}$ as the domain of the proximity relation $P_\Sigma^{\alpha+1}$, the latter whose existence follows from the Bisimulation Principle. We thus have a proximity space $(b^{\alpha+1}, P_\Sigma^{\alpha+1})$. Finally, we make $(b^{\alpha+1}, P_\Sigma^{\alpha + 1})$ into the Kripke structure $ \kri_\Sigma^{\alpha+1}$ by assigning to each node $\psi_i \in b^{\alpha+1}$  the truth value $v_j(\psi_i) = 1$ for $i = j$, with $1 \leq i,j \leq N'$. This completes both the proof of Proposition \ref{bpresults} and our inductive proof stated at the beginning of section \ref{preferred bases}.

\section{Conclusions and outlook \label{conclusions}}

Let us return to the five problems formulated in the introduction. With the Bisimulation Principle and with the Imperfection Principle we have attempted a plausible explanation to all five. The Bisimulation Principle can thus be seen as a {\it meta-concept} providing a new perspective on these problems and on the two elementary ingredients which we have used: the structural unfolding process of analytical set theory, and the proximity relation acting on all possible experiment outcomes. We have argued that the former admits a natural mathematical form of the stage paradigm in the quantum universe; while, on the other hand, the latter is a direct mathematical expression of the experimetalist's fundamental inability to perform perfect experiments. Here, the Imperfection Principle is seen as an explicit incorporation of Poincar\'e's physical continuum into a quantum theory of the universe, and proximity spaces, which provide a general approach to quantum logic, are the mathematical input for the Imperfection Principle. Both ingredients are formulated in quite simple mathematical terms, but at the same time they provide some interesting mathematical and physical implications. Not only a possible solution to the preferred basis and the state selection problems in the context of a self-referential quantum universe (that is, a quantum universe formulated on the basis of non-wellfounded objects), but also a generalization of Born's rule due to Dempster-Shafer theory, a connection between tree metrics and binary codes, and a characterization of the quantum universe in terms of tree metrics. Moreover, we have argued that the natural appearance of tree metrics might be a hint that there exist other fundamental degrees of freedom of the universe {\it beside} pure geometrical degrees, like positions in three-space. This characterization of the quantum universe suggests that information sequentially encoded on physical carriers might not only be essential in evolutionary biology but that it could also play a decisive role in {\it evolutionary cosmology}.\medskip

Returning to the problem of how a smooth manifold structure of the universe could be recovered from this approach, we refer briefly to {\it phylogenetic algebraic geometry} \citep{eri2004}. In this young mathematical field complex algebraic varieties are studied that are naturally associated with probabilistic phylogenetic trees. These are tree structures where nodes and leafs represent finite probability distributions and where arcs between nodes represent probabilistic transition matrices \citep{eri2004}. Real positive points on these complex projective varieties represent probabilistic models of evolution and several examples of well-known classical algebraic varieties have been found: toric and among them Veronese and Segre varieties, and as well secant varieties (these appear for other tree topologies that do contain hidden nodes). So far, all examples of phylogenetic algebraic varieties have been found through analysis of small trees with only a few nodes or leafs, and algebraic geometry for large (finite) phylogenetic trees is still unexplored territory. Here an interesting connection of the present work to phylogenetic algebraic geometry appears since our results in section \ref{metric spaces} and particularly Corollary \ref{coll1} tell that each Kripke structure $\kri_\Sigma^\alpha$ can be identified as a {\it probabilistic} phylogenetic tree in the above sense (that is, every node or leaf of  $\kri_\Sigma^\alpha$ can be represented as an element of a non negative measure space $({\cal A}_\mu, d_\mu) \subset L_1(\Omega,{\cal A}, \mu)$). Although not yet firmly established, we think that this connection to phylogenetic algebraic geometry allows for a global and a local view on $\kri_\Sigma^\alpha$. In taking on the global view one would try to study the algebraic variety of each $\kri_\Sigma^\alpha$ in total, that is, one would try to find {\it one} variety of immense complexity (by means of its number of dimensions) representing $\kri_\Sigma^\alpha$. On the other hand, the local and more familiar view on the quantum universe could reveal {\it many} local algebraic varieties of low dimension attached to each branching point in $\kri_\Sigma^\alpha$. This seems not implausible because $\kri_\Sigma^\alpha$ is a nested phylogenetic tree structure consisting of many small trees that altogether form the large tree. One goal of such a local approach would be the representation of the quantum universe $\kri_\Sigma^\alpha$ as a smooth algebraic variety, i.e. as an algebraic manifold, by looking for ways to define a natural continuous transition between neighboring local varieties based on $P$-continuity and its induced topology.\smallskip 

Our final remark is about the evolutionary rules of the quantum universe. As we have argued these rules include three initial entities: the unknown universe $U$, the structural unfolding rule in the form of equation (\ref{unfoldrule}), and the initial modal sentence $\varphi_U^0$. These initial conditions determine the unfolding process. Is there possibly a freedom of choice in the quantum universe, for example, a freedom that permits to {\it overwrite} these rules with new ones? Under the assumptions that the {\it unknown set} $U$ is inaccessible to any changes and that the unfolding rule is of universal validity, a possible way to alter these rules would be a {\it semantic change} in the initial modal sentence $\varphi_U^0 \mapsto \bar\varphi_U^0$. Buccheri argues that {\it rules of rules} may exist within a quantum universe when formulated on the stage paradigm \citep{buc2003}, and he conjectures that such rules should have probabilistic mathematical form. Future work could therefore look for certain probabilistic or statistical models that represent such overwriting of rules.\bigskip\bigskip

Acknowledgments. The author thanks the Bay\-erische Staats\-bib\-liothek M\"unchen for hospitality during the preparation of this work.

\vfill
\addcontentsline{toc}{section}{References}
\bibliographystyle{plain}
\bibliography{modal_v3}

\end{document}